\documentclass[sigconf]{acmart}
\usepackage{array}
\usepackage{multirow}
\usepackage{xcolor}
\usepackage{graphicx}
\usepackage{caption}
\usepackage{comment}
\usepackage{color,soul}
\usepackage{lipsum}
\usepackage{hyperref}

\AtBeginDocument{%
  \providecommand\BibTeX{{%
    \normalfont B\kern-0.5em{\scshape i\kern-0.25em b}\kern-0.8em\TeX}}}

\copyrightyear{2023} 
\acmYear{2023} 
\setcopyright{rightsretained} 
\acmConference[CHI '23]{Proceedings of the 2023 CHI Conference on Human Factors in Computing Systems}{April 23--28, 2023}{Hamburg, Germany}
\acmBooktitle{Proceedings of the 2023 CHI Conference on Human Factors in Computing Systems (CHI '23), April 23--28, 2023, Hamburg, Germany}\acmDOI{10.1145/3544548.3580718}
\acmISBN{978-1-4503-9421-5/23/04}

\begin{document}


\title[Using Virtual Reality to Shape Humanity's Return to the Moon]{Using Virtual Reality to Shape Humanity's Return to the Moon: Key Takeaways from a Design Study}

\author{Tommy Nilsson} 
\affiliation{ 
    \institution{European Space Agency (ESA)}
    \city{Cologne}
    \country{Germany}} 
\email{tommy.nilsson@esa.int}

\author{Flavie Rometsch }
\affiliation{ 
    \institution{European Space Agency (ESA)}
    \city{Noordwijk}
    \country{Netherlands}} 
\email{flavie.rometsch@esa.int}

\author{Leonie Becker} 
\affiliation{ 
    \institution{German Aerospace Center (DLR)}
    \city{Braunschweig}
    \country{Germany}} 
\email{leonie.becker@dlr.de}

\author{Florian Dufresne} 
\affiliation{ 
    \institution{Arts et Métiers Institute of Technology}
    \city{Laval}
    \country{France}} 
\email{florian.dufresne@ensam.eu}

\author{Paul de Medeiros}
\affiliation{ 
    \institution{European Space Agency (ESA)}
    \city{Cologne}
    \country{Germany}} 
\email{hello@pauldemedeiros.nl}

\author{Enrico Guerra} 
\affiliation{ 
    \institution{European Space Agency (ESA)}
    \city{Cologne}
    \country{Germany}} 
\email{enricoguerra@outlook.com}

\author{Andrea E. M. Casini}
\affiliation{ 
    \institution{German Aerospace Center (DLR)}
    \city{Cologne}
    \country{Germany}} 
\email{andrea.casini@dlr.de}

\author{Anna Vock}
\affiliation{ 
    \institution{Politecnico di Milano}
    \city{Milano}
    \country{Italy}} 
\email{annalea.vock@mail.polimi.it}

\author{Florian Gaeremynck}
\affiliation{ 
    \institution{European Space Agency (ESA)}
    \city{Cologne}
    \country{Germany}} 
\email{florian.gaeremynck@gmail.com}

\author{Aidan Cowley} 
\affiliation{ 
    \institution{European Space Agency (ESA)}
    \city{Cologne}
    \country{Germany}} 
\email{aidan.cowley@esa.int}

\renewcommand{\shortauthors}{}

\begin{abstract}

Revived interest in lunar exploration is heralding a new generation of design solutions in support of human operations on the Moon. While space system design has traditionally been guided by prototype deployments in analogue studies, the resource-intensive nature of this approach has largely precluded application of proficient user-centered design (UCD) methods from human-computer interaction (HCI). This paper explores possible use of Virtual Reality (VR) to simulate analogue studies in lab settings and thereby bring to bear UCD in this otherwise engineering-dominated field. Drawing on the ongoing development of the European Large Logistics Lander, we have recreated a prospective lunar operational scenario in VR and evaluated it with a group of astronauts and space experts (n=20). Our qualitative findings demonstrate the efficacy of VR in facilitating UCD, enabling efficient contextual inquiries and improving project team coordination. We conclude by proposing future directions to further exploit VR in lunar systems design.

\end{abstract}

\ccsdesc[500]{Human-centered computing~Virtual reality}
\ccsdesc[500]{Human-centered computing~Human computer interaction (HCI)}




\keywords{Virtual Reality, human factors, ergonomics, user centered design, HCI research, space system engineering, lunar lander}


\maketitle

\section{Introduction}
50 years after Apollo astronauts last set foot on the Moon, an unprecedented international collaboration seeks to send humans back to the lunar surface as part of the NASA-led Artemis programme. With a first crewed landing planned for 2025, the long-term goals include establishing permanent lunar bases before the end of the decade \cite{Smith2020}. Surpassing Apollo in scope and ambition, the success of humanity’s renewed push for Moon will rest on reliable, safe and effective technologies being designed in support of future crews and their extravehicular activities (EVA’s). \textcolor{black}{Accordingly, the development of novel methods for advancing low-technology readiness level (TRL) solutions in service of human and robotic operations on the Moon has been described as a key ambition of the Artemis programme \cite{Weber2021}. }

Designing such solutions requires taking into account challenges and limitations posed by a range of environmental, operational and human factors that may markedly deviate from established terrestrial usability and ergonomics considerations. This includes elements such as the reduced lunar gravity, dust mitigation, limited field of view and range of motion due to EVA spacesuits, mental and physical fatigue, thermal and atmospheric situation, limited capability for redundancy and the extreme lighting conditions, such as pitch black shadows \cite{eppler1991lighting, Moore1989}, all against the background of often ambiguous mission criteria and still largely undefined operational scenarios \cite{beaton2019b}.

The lighting conditions in particular have emerged as a prominent consideration during design and preparatory activities. Notably, the tilt of the lunar axis creates a unique setting on the lunar south pole, with elevated areas experiencing near-permanent illumination, whilst the very low sun angles simultaneously also block out all sunlight in depressed areas. Indeed, the access to persistent illumination (and the resulting availability of solar power), along with water ice preserved in permanently shadowed areas, was a key factor leading to the selection of the Moon’s south polar region as the likely site of the next human landing and the subsequent construction of base camps \cite{Smith2020}.

In a bid to factor in such lunar conditions during design and development processes, engineering teams have typically resorted to experimental deployments and field studies of prototypes in analogue environments, such as large cave systems to approximate the poor lighting conditions on the Moon \cite{sauro2023training}, or neutral buoyancy pools to simulate lunar gravity \cite{Bessone2015}.

These \textit{classical} approaches have met with varying degrees of success, attracting criticism for their logistical complexity and high costs, typically resulting in low frequency of experimental deployments and limited number of test subjects \cite{casini2020lunar}. 
\textcolor{black}{Barriers like these have prevented widespread adoption of established UCD methodologies, such as rapid prototyping and participatory design, that are otherwise successfully employed in HCI. As a result, space systems development projects are often plagued by limited agility, frequent delays and notorious budget overruns \cite{Foust2021}. }

In response, this paper investigates potential use of Virtual Reality to simulate field studies and thus facilitate rapid and resource-efficient user-centered assessments of early-stage lunar surface prototypes. Although VR’s capacity to interactively simulate hypothetical concepts and environments at comparatively low costs has already garnered significant attention in several fields \cite{walch2017evaluating, garassini1994evening}, its applicability in simulating the unique lunar conditions and facilitating assessments of pertinent design solutions remains unexplored. 

The goal of our work, then, is to understand the strengths and weaknesses of VR-based field studies in terms of their capacity to enable user-centered HCI methods in the context of lunar system design. By shedding light on the types of design questions this approach can and cannot reasonably address, we seek to lay a foundation for responsible use of VR to ameliorate ongoing and future design projects in this domain.

To this end, we created an interactive simulation of the Moon’s south polar region in VR for the purpose of assessing a design concept of the European Large Logistics Lander (EL3); a prospective autonomous lunar landing vehicle currently under development by an international consortium led by the European Space Agency (ESA) \cite{Duvet}. Working closely with the EL3 development teams, we produced a plausible operational scenario and evaluated it with human spaceflight experts, including astronauts, engineers, astronaut instructors, project managers and scientists.

Their qualitative reflections have demonstrated the utility of VR in enabling user-centered approaches to studies of operational performance and human factors in the early stages of a lunar surface system’s design and development, thus significantly expanding the potential applicability of UCD methodologies in this traditionally engineering-dominated discipline. By highlighting systemic aspects of future lunar operations, we found VR-based scenarios to be uniquely qualified for driving reflection on the synergies and frictions within broader human-machine ecosystems. Furthermore, the comparatively accessible nature of VR proved well suited to coordinate and improve the agility of project teams by virtue of facilitating a model-based systems engineering approach. By considering these findings, we take a first step towards understanding the potential use of VR as a design tool to shape humanity’s expansion to the Moon and beyond.

\textcolor{black}{
\section{
Statement of Contribution
}
The contribution of our work is threefold: (1) We introduce VR-simulated field studies as a novel methodology in space system design. To determine its efficacy, we present, to the best of our knowledge, the first documented application of a UCD approach during the development of a concept-stage lunar surface system. (2) We extend the ongoing discourse on immersive design scenarios and speculative enactments by reflecting on their potential use in terms of facilitating contextual and holistic assessments of human-machine ecosystems as well as supporting the coordination of design project teams. (3) We provide a set of recommendations on VR interfaces used to simulate lunar conditions (e.g. low gravity), including possible ways of leveraging mixed reality technology to improve the validity of future design inquiries.} 

\textcolor{black}{
\section{Related Work}
There is a wide range of HCI and VR publications related to our work. Here, we focus on three main topics: (1) HCI research methods in space systems design, (2) classical approaches to assessing space design concepts via analogue field studies, and (3) the potential use of VR to bridge the gulf between these two. }
 
\textcolor{black}{
\subsection{HCI in Space Systems Design}
Aerospace engineering and HCI have been intertwined since the dawn of the space age. Efforts in this vein have materialized into solutions spanning from computer interfaces supporting physical and mental wellbeing of astronauts \cite{pataranutaporn2021spacechi} to telerobotic control mechanisms \cite{sheridan1992telerobotics}. Notably, influential ideas from HCI, such as \textit{human-machine symbiosis} \cite{clynes1960cyborgs}, were leveraged throughout the development of modern spacesuit designs to enhance human performance and safety during EVA operations \cite{de2011spacesuit} and helped shape numerous relevant solutions, including exoskeletons for supporting humans in low-gravity \cite{porter2020soft, sumini2020spacehuman} or augmented reality interfaces for improved spatial orientation \cite{helin2018user}.  }

\textcolor{black}{
Nevertheless, the development of such systems has traditionally been engineering-driven, guided predominantly by utilitarian metrics, such as technological feasibility and crew safety. In contrast, comparatively little attention has been paid to the use of HCI design methods\footnote{We base our use of this term on the work of de Haan, who defined HCI design methods as \textit{“human-centred and creative approaches to the conceptualising and building of user interfaces”} \cite{de2015hci}.} more directly by involving end users and other stakeholders in the design process \cite{trotta2020communicating}.  
}

\textcolor{black}{
The ongoing proliferation of \textit{New Space companies} \cite{weinzierl2018space}, such as SpaceX, and the consequent democratization of access to space, has therefore led some scholars to suggest that the time has come for HCI to play a more central role in development of future space systems\cite{pataranutaporn2021spacechi}. As elaborated by Trotta et al., the HCI community now has a unique opportunity to \textit{“contribute a new perspective and knowledge on how to think about and design future interfaces in space”} \cite{trotta2020communicating}.  
}

\textcolor{black}{
Early efforts to take up this challenge have typically resulted in various forms of UCD approaches employed for the purpose of engaging non-engineers in ideation of future space solutions. For example, in a series of CHI workshops, Pataranutaporn et al. invited scholars from across disciplines to brainstorm around the possibilities offered by HCI for space design. The result was a wide range of novel scenarios and use cases for space HCI research centered around emerging stakeholders, such as space tourists \cite{pataranutaporn2021spacechi, pataranutaporn2022spacechi}. Similar approaches have been utilized to conceptualize hypothetical lunar settlements \cite{casini2018analysis, vock2022holistic}, or even to co-create new ideas around the future of food and eating in space \cite{obrist2019space}.  
Closer to our work, Rometsch et al. conducted a series of group brainstorm sessions with aerospace experts to converge on potential cargo unloading solutions for future supply shipments to the lunar surface \cite{rometsch2022towards}. 
}

\textcolor{black}{
However, apart from ideation, the application of UCD during prototyping and evaluation of real-world solutions has been sparse. While examples exist of software tools for human space exploration being developed in a user-centric manner \cite{boy2013situation, platt2014participatory}, applying such approaches during the design of physical interfaces has proven to be more complicated. For instance, Sumini et al. developed a seahorse-inspired prosthetic tail designed to enhance body motion in low-gravity conditions \cite{sumini2020spacehuman}. To evaluate its effectiveness, assessments had to be carried out on board reduced-gravity flights. In this case, the relatively low prototype development costs meant the authors could arrange a series of such prototypical deployments, allowing for a user-centered iterative design process. }

\textcolor{black}{
In contrast, many space agencies currently engaged in development of comparatively more complex solutions for the Moon, such as lunar landers or rovers, have to deal with greater constraints. As will be detailed in the following section, the high financial and logistical burden of assessments, combined with the need to evaluate such solutions in a representative environmental and operational context, have largely hampered the adoption of HCI methodologies (and UCD in particular) in classical approaches to lunar systems development.  }

\subsection{Classical Field Studies}
It is common practice that novel lunar and planetary exploration solutions are tested and evaluated in so-called \textit{analogue environments}. Several sites, such as the caves of Sardinia \cite{Bessone2015} or the volcanic sand plains of Iceland \cite{gasser2019} share important characteristics with the lunar environment, including terrain features or geological composition. Aside from such naturally occurring analogues, some lunar conditions have also been simulated by artificial means. For example, reduced-gravity flights and neutral buoyancy pools have been employed to replicate the Moon’s partial gravity levels (0.17g), while vacuum chambers have been used to simulate the lack of lunar atmosphere \cite{Dorota}.

\textcolor{black}{In addition to simulating relevant environmental conditions, another core purpose of analogue studies is the assessment of design solutions within the context of prospective mission scenarios \cite{beaton2019b}. This typically entails the study of their synergies with other key operational concepts and elements of the mission architecture, including interhuman communication, data collection and distribution systems, scientific procedures, logistical workflows and remote supervision \cite{Turchi2021}.}
 
Space agencies are increasingly turning to such analogue field trials to inform the design of reliable solutions for lunar exploration \cite{Beaton2020, Osinski2006, Leveille2010}. For example, in 2019 ESA conducted a field campaign in the arid volcanic landscape of Lanzarote in the Canary Islands to evaluate several prototypical tools for the collection and storage of geological samples on the Moon’s surface \cite{Pangaea-X}. One of the tests, for instance, saw a team of experts compare the functionality and maneuverability of two alternative trolleys, seeking to find the optimal balance of properties \cite{NEST}.

\begin{figure*}[htp]
    \centering
    \captionsetup{justification=centering}
    \includegraphics[width=\linewidth]{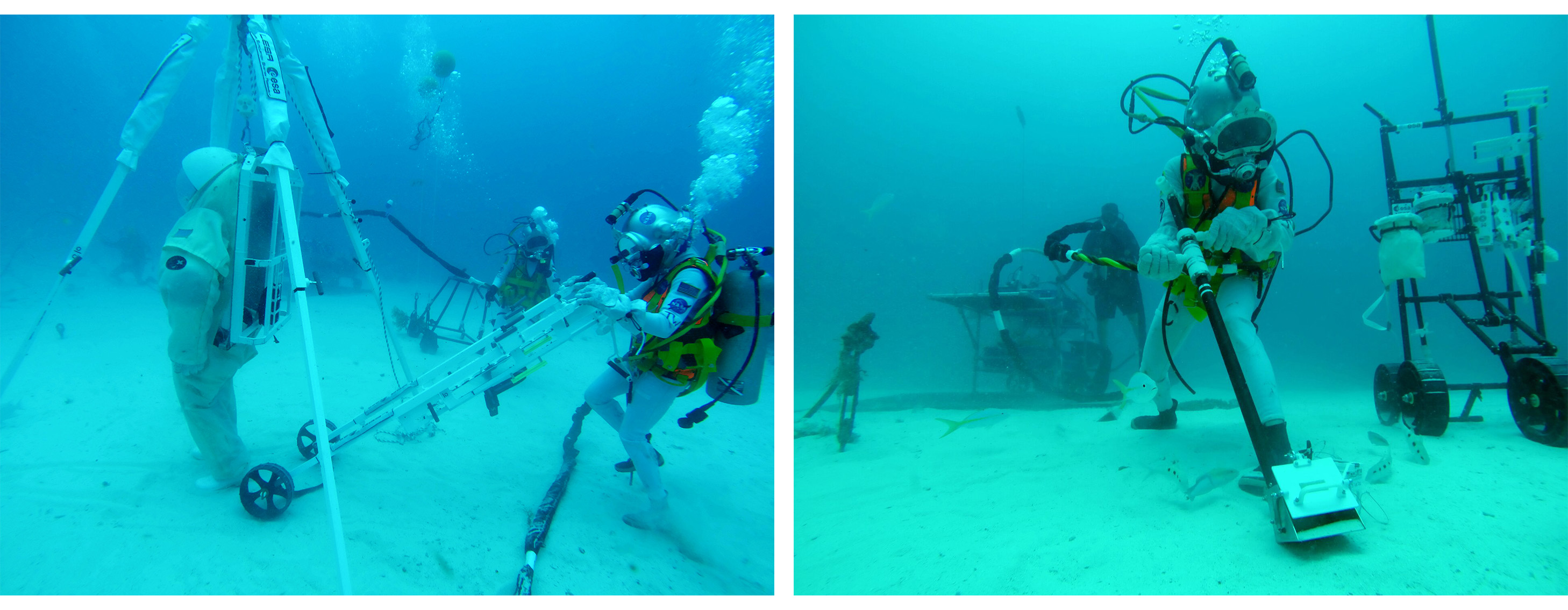}
    \caption{An astronaut simulating the rescue of an incapacitated crew-member with a portable vehicle designed for evacuation in lunar gravity. The underwater weight of this suit simulator was equivalent to the weight of an astronaut wearing an EVA suit on the Moon \cite{LESA} (left). An astronaut performing simulated EVA operations during an analogue study \cite{NEST} (right).}
    \label{fig:classicalApproaches}
    \Description[Astronauts evaluating design prototypes during analogue studies]{Analogue studies have been conducted on the Atlantic Ocean seabed in order to evaluate the performance of design prototypes in a reduced gravity environment.}
\end{figure*}

In a separate 2019 study, two astronauts performed a series of underwater spacewalks in the Atlantic Ocean with the aim of evaluating a prototypical portable vehicle designed for the evacuation of incapacitated astronauts under lunar gravity conditions \cite{equipment} (see Figure \ref{fig:classicalApproaches}). Future crew members were thus given the opportunity to reflect on the (inter)operability and ergonomics of relevant systems, as well as potential risks associated with their use. This included ergonomic and mechanical aspects surrounding interactions with relevant components such as pulley systems, handles, locomotion systems and other tool interfaces while being weighted-out to the lunar gravity, wearing mock-up EVA gloves and a representative helmet reproducing a spacesuit’s constrained field-of-view \cite{LESA}, all against the backdrop of the uneven sandy and rocky seabed \cite{Dorota}. 

Although the feedback gathered through such studies has helped facilitate iterative integrated design and development processes for potential lunar surface solutions, the reliance on analogue environments is not without its drawbacks. Mainly, the choice is often limited to single-aspect environments with only a few analogues capable of replicating multiple lunar conditions simultaneously (e.g. only geology and gravity, but not the lighting conditions in the case of seabed trials \cite{Dorota}). Because one analogue cannot simulate all the different aspects of a lunar mission, a variety of analogue and artificial testbeds is typically required. In addition, these analogue campaigns require significant preparation and planning time, consuming large amounts of resources. They also generally only allow a small number of participants to take part in the tests, limiting the volume of design assessments that can realistically be carried out on a yearly basis.

Such limitations have traditionally predisposed relevant design projects to adopting the relatively rigid \textit{V-model} approach, with considerable amount of development taking place prior to any experimental deployments and evaluation through analogue testing activities \cite{Dorota}. \textcolor{black}{In contrast to typical design projects in HCI, the interaction between development teams and end users in space systems engineering is thus rather limited and often only occurs towards the end of a development cycle. As a result, by the time an operational scenario is deployed and assessed at the end of the V-model design cycle, there tends to be little room left to influence the design. This results in longer feedback loops and slower design iterations, incurring further financial and temporal costs on the project.} 

\textcolor{black}{
In the following section we will argue that some of these limitations might be alleviated through greater use of VR-based design practices during early stages of a project's lifecycle.}

\subsection{Design Studies in Virtual Reality}
The aerospace industry was at the forefront of popularizing VR technology. Notably, NASA pioneered the use of this tool when training the Hubble space telescope flight team for a repair mission in 1993 \cite{loftin1995training}. The benefits of using VR in this case were evident, as it reduced EVA training time and therefore project cost.

However, it was not until the 21st century that VR and virtualization began to gradually gain a foothold in design practices surrounding space systems development. Virtual environments have since been utilized, for instance, to help prevent misalignments and erroneous data flow in the International Space Station’s Columbus module \cite{cardano2009vr}, to assist with general satellite assembly troubleshooting \cite{geng2017virtual} or to analyze the replacement procedure of a Columbus cabin filter \cite{helin2017augmented}.

VR technology has also seen use during verification and validation project phases, such as assembly scheme and maintenance reviews for the Orion lunar spacecraft \cite{via_satellite_2017} or to collect astronaut feedback for an ergonomic evaluation of the Gateway lunar space station modules \cite{thales1}. Nevertheless, such cases are still rare, with VR usually taking a back seat to analogue field studies when more systematic user evaluations are required.

Instead, it is non-space related domains that are now leading the way in the adoption of VR-centeric design processes. Indeed, VR has already been successfully used to study human interaction with a range of prototypical technologies, including autonomous cars \cite{sportillo2017immersive}, healthcare solutions \cite{lohse2014virtual} and complex robotic systems \cite{miner1994interactive}. Such studies have demonstrated the ability of VR simulations to gather a wide range of relevant user feedback on topics including usability \cite{drey2020vrsketchin}, human factors and ergonomics \cite{wienrich2018assessing}, safety \cite{somin2021breachmob} and acceptability \cite{fussell2021using}. By collecting such feedback in early stages of a project lifecycle, development teams have been able to better anticipate potential problems and lower risks for the overall production processes, resulting in a reduction in development and manufacturing costs \cite{geng2017virtual, jerald2015vr, berg2017industry}.

\textcolor{black}{
Compared to studies centered around traditional physical prototypes, VR was found to enable rapid iteration of design concepts without being limited by their number, size, or shape, nor the spatial constraints of the lab \cite{jetter2020vr}. Similarly, evidence suggests that VR is superior to digital mockups experienced in 2D when it comes to facilitating spatial assessments \cite{hubenschmid2022relive} and stimulating creative and flexible thinking \cite{lee2019design}.}

Another aspect of VR that has recently attracted attention of the HCI community is its capacity to embed design concepts in a realistic context. For instance, Mäkelä et al. evaluated the use of information displays in public spaces by simulating their placement in a representative virtual environment. They found that by providing users with a context that simulates a natural scenario, VR contributes to a more realistic experience than what could otherwise have been achieved in a lab setting \cite{makela2020virtual}. Drawing on established HCI methods, such as scenario-based design or design fiction, Simeone et al. follow a similar line of reasoning, arguing that extending such “speculative enactments” to Virtual Reality would enable users to engage with design ideas in their \textit{"truest envisioned form"} \cite{simeone2022immersive}. Based on their exploration of immersive design fictions, McVeigh et al. have echoed similar views, arguing that VR is particularly well suited for prototyping embodied and contextually rich aspects of speculative experiences \cite{mcveigh2018immersive}.

\textcolor{black}{
Given its ability to immerse prospective users in such contextually rich experiences, coupled with its applicability in early design stages, we could posit that VR is uniquely qualified to support the development of future lunar surface solutions. Nevertheless, VR-enabled design processes in space system development still have a long way to go before they can be considered a standardized approach. There is an opportunity, then, to come full circle by enabling human spaceflight to fully capitalize on VR technology it once helped pioneer. In return, VR might open up this traditionally engineering-dominated field to greater application of user-centric design methods, along with all the aforementioned benefits that come with such approaches.}

\textcolor{black}{
By producing a VR-based scenario centered around a prospective lunar lander and evaluating it with expert users, as described in the following section, we hope to shed further light on the viability and validity of VR as a design tool in this unique domain. }   

\section{Methodology}
\begin{figure*}[hbt!]
    \centering
    \includegraphics[width=\linewidth]{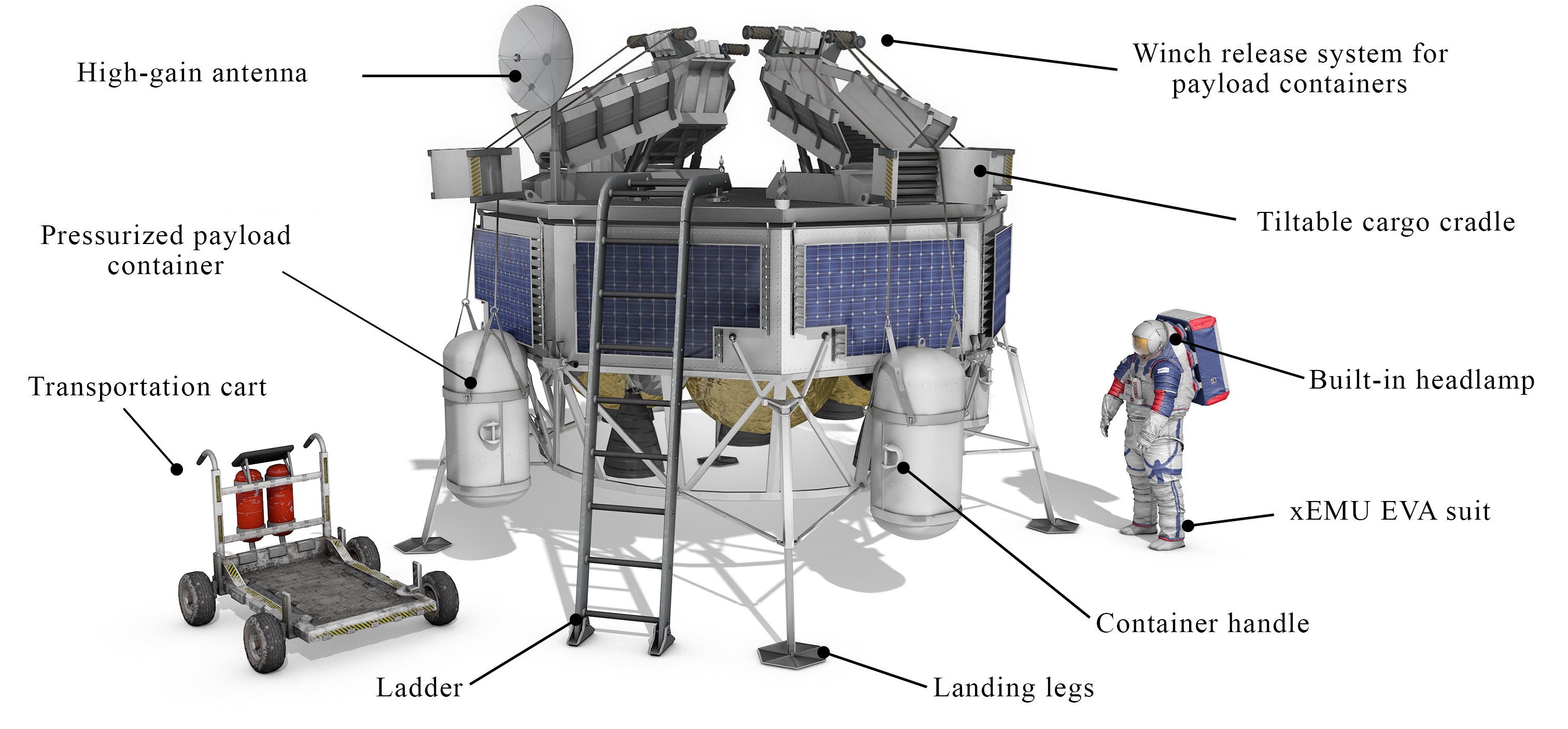}
    \caption{3D model of our agnostic configuration of the EL3 lander, along with relevant assets}
    \label{fig:assets}
        \Description[An overview of VR assets included in our study]{Aside from the EL3 lander, our VR study featured a range of other prospective lunar solutions, including the xEMU spacesuit, cargo containers and a transportation cart.}
\end{figure*}

In order to explore the potential use of VR to catalyze UCD approaches in the context of human lunar exploration, we carried out a simulated field study focused on a VR-based operational lunar scenario. Specifically, we created a virtual mockup depicting a prospective configuration of the EL3 lunar lander along with a potential cargo unloading system. This was complemented with 3D models of cargo containers, a transport cart, an EVA space suit and a ladder to climb the lander (see figure \ref{fig:assets}). A virtual replica of the Moon's south polar region was produced to form a realistic backdrop. These assets were then combined into an interactive simulation of cargo-reception and offloading procedures on the lunar surface. Finally, borrowing from established HCI methods, such as Scenario-Based Design \cite{carroll2003making}, we invited a group of relevant experts to immerse themselves into this VR scenario and critically evaluate the simulated design solutions. In the remainder of this section, we will elaborate on these steps in more detail. 

\subsection{The EL3}
The feasibility of a sustained human presence on the Moon will depend heavily on the development of reliable logistical solutions for delivery of crew supplies and other forms of cargo to the lunar surface. Recent years have seen a number of both public and private stakeholders from around the world taking up this challenge, spawning a spectrum of design concepts ranging from the 325 kg light Astrobotic Peregrine Lander \cite{chavers2016nasa} to the towering 100 ton SpaceX Starship Human Landing System \cite{seedhouse2022starship}. 

In line with this trend, ESA, together with its partners from the industry and academia, is currently in the process of designing the European Large Logistics Lander (EL3), an autonomous lunar landing vehicle capable of delivering a wide range of crew supply payloads to the Moon, with an initial launch window scheduled between 2028 and 2029. Once taken into service, the EL3 is expected to form the backbone of Europe’s pathway toward sustainable human exploration of the Moon \cite{gollins2020, Carey2021, landgraf2022autonomous}.  

Upon completion of the initial mission requirements specification phase and securing of funding, which is expected to take place by the end of 2022, the industrial implementation phase will be initiated \cite{Duvet}. The first design reviews of the EL3 are then scheduled to commence in the first quarter of 2023 \cite{Carey2021}.

The important role the EL3 is expected to play in future human surface operations on the Moon, in combination with its still-untested design, made it an ideal subject for our study.

Guided by input from the EL3 project management and responsible engineering team, we modeled an agnostic EL3 configuration representing a generic design that could meet the requirements outlined in the EL3’s original invitation to tender \cite{Carey2021}. The resulting EL3 model is approximately 2.8 m tall, with an approximately 14 m\textsuperscript{2} large octagonal cargo deck on top. We have attempted to provide the model with the highest possible level of detail to allow for the evaluation and study of issues related to overall usability, operations and human factors challenges and constraints, which are likely to apply to most potential designs that might result from the EL3 team’s own work.

On top of the virtual lander, we designed and implemented a hypothetical cargo unloading system in the form of a winch-based pulley system with cradles that would rely on lunar gravity to deploy cargo containers from the deck at the top. The cargo unloading mechanism was animated in the VR scenario. This concept was once again guided by input from the EL3 development teams, making sure our virtual mockup was reasonably realistic. Further details concerning this process can be found in our precursor study \cite{nilsson2022using}.

\subsection{Additional Assets}
A number of additional 3D assets required for the scenario was produced based on real-world references. 

Firstly, the EL3 is being designed to carry a set of specific cylindrically shaped pressurized cargo containers. These are expected to have a mass of ca. 350 kg and a pressurized volume of ca. 0.7 m\textsuperscript{3}. 4 of these were reproduced virtually and served as essential components of our  scenario, as participants were tasked to retrieve these containers and carry them. This meant that the cargo containers had to be physics components and be interactable in VR.

Secondly, to enable users to be able to inspect the cargo unloading system from the top of the lander and to elicit potentially important insights, as well as to help predict potential scenarios (e.g., required maintenance due to a jammed system), a climbable ladder was designed and integrated. The VR user was able to grab the rungs and/or side rails of the ladder by latching onto these with their VR hands using the motion controllers and move up or down the ladder respectively, simulating climbing.

Thirdly, a static cart was added to trigger valid reflections on potential means of container transportation.

To further improve authenticity of the scenario, users were embodied in a 3D model of the a spacesuit. Specifically, we employed the Exploration Extravehicular Mobility Unit (xEMU) EVA suit. Users had their heads encapsulated inside the xEMU helmet to accurately restrict their field of view \cite{ross2018nasa}. The xEMU helmet was likewise equipped with a headlamp. It should be noted that only the torso and suit gloves were visible to the user, legs and boots were not included, nevertheless the user was casting a full body shadow.

All of the 3D models were textured using physically based rendering materials to accurately mimic light interactions, such as glare and reflections.

All assets were simulated purely visually in our scenario. Complementing visual VR with physical mockups of the simulated objects was briefly considered, but eventually abandoned due the technical difficulties involved in making physical mockups follow the laws of the reduced lunar gravity. We shall reflect on the limitations stemming from this approach in the closing discussion of this paper.

\subsection{Virtual Moonscape}
Based on topographic maps captured by the Lunar Reconnaissance Orbiter \cite{smith2017summary}, we recreated virtually an area of 64 km\textsuperscript{2} in close vicinity of the Shackleton crater on the Lunar south pole (89.9°S 0.0°E). We selected this area as it has been identified as one of the candidate landing sites for the first Artemis human landing mission \cite{Smith2020}, thus providing a sufficiently plausible backdrop to our scenario. Due to the original topographic map‘s pixel scale of 100 meters, smaller details (e.g. boulders) were added manually using 3D modeling software. The virtual moonscape was subsequently textured procedurally using a set of custom created terrain shaders. Particle effects were implemented to visually mimic the plumes of moondust generated by physical contact with the lunar surface. A lunar geologist from our organization assisted us during this process to ensure authenticity.

\subsection{The Scenario}
Finally, the virtual environment was instantiated using the Unreal Engine 4 game engine. The sun was placed in the direction of north, at an angle of 1.5° above the horizon and its intensity set to 1.37 kW/m\textsuperscript{2} to mimic the conditions on the lunar south pole \cite{vanoutryve2010analysis}. All forms of indirect lighting and light scattering were disabled to recreate the pitch-black shadows stemming from the lack of lunar atmosphere.

The EL3 mockup carrying four cargo containers was placed in the middle of our virtual moonscape. A cargo drop-off point was placed 30 meters away from the lander and marked by a flag planted in the ground. Our scenario thus conveyed a basic, yet plausible, situation with users being tasked to approach the lander, retrieve the containers, and transport these to the nearby drop-off point. The ladder and transportation cart were included in the scene to trigger discussion on potential operational scenarios.

The VR experience was run on a desktop computer with an RTX 2080 GPU, providing smooth framerate. We used the HTC Vive Pro VR headset coupled with a pair of HTC Vive base stations and controllers. Walking and turning was handled via the controller trackpads, while the controller triggers were used for interaction with objects in the virtual environment (e.g. lifting up payload containers). Users could also look around the environment by moving their head. This VR cargo unloading scenario then formed the focal point of our engagement with participants.

\begin{table*}[hbt!]
\resizebox{\textwidth}{!}{%
\begin{tabular}{|l|c|l|l|}
\hline
\textbf{Participant} & \textbf{Gender} & \textbf{Job Title}                              & \textbf{Area of Expertise}                             \\ \hline
Astronaut 1                 & M               & Astronaut                                       & Human Spaceflight,   EVA                               \\
Astronaut 2     & M               & Astronaut                                       & Human Spaceflight, EVA                                 \\

Engineer 1           & M               & ISS Operations Engineer                         & Human Spaceflight Operations                           \\
Engineer 2           & M               & Crew Technologies and Astronaut training     & Astronaut Robotics Training, XR Technologies         \\
Engineer 3        & F               & Ops Engineer                            & Aerospace engineering                  \\
Engineer 4           & M               & ISS Ground Segment Engineer                     & Human Spaceflight Operations, Avionics                           \\
Engineer 5              & M               & ISS Ground Segment Engineer                         & Human Spaceflight engineering, VR                                     \\

Instructor 1         & M               & Astronaut Instructor                            & ISS payloads specialist, XR training                                     \\
Instructor 2      & M               & Astronaut Instructor / ODF (Operations Data File) manager              & Crew training and Operations                                     \\
Instructor 3        & F               & Columbus Instructor and EUROCOM                 & Astronaut Training, Human Factors                                     \\
Instructor 4       & M               & Astronaut and Ground Personnel Instructor       & Human Spaceflight Operations                                     \\

Instructor 5                   & M               & Astronaut Instructor and Ops Engineer           & Payload training and EUROCOM  \\

Instructor 6      & M               & ISS Astronaut Instructor and EUROCOM            & Columbus Ops, astrophysics                                    \\
Instructor 7   & M               & Astronaut Instructor                            & ISS experiment training and operations                                     \\

Instructor 8                 & F               & Astronaut Instructor                            & Astronaut Training, Aerospace engineering                                     \\ 
Manager 1                & M               & Manager of an EVA and Parabolic Flight Training Unit & EVA Training for Moon and Mars, Analogue simulations                                           \\
Manager 2              & M               & Science and Operations Manager                  & Human Spaceflight, Lunar Science and XR   Technologies \\

Scientist 1                 & M               & ISS Operations and Training Support Scientist & Astrophysics, planetary science                    \\
Scientist 2              & M               & Research Fellow                                 & Instrumentation development for astronaut analogue  training                    \\
Scientist 3            & M               & Science Advisor                                 & Human Spaceflight and Lunar Science                    \\
\hline
\end{tabular}%
}
\caption{An overview of the study participants including gender, title and (self-reported) areas of expertise}
\label{tab:participants}
    \Description[Study participants]{20 expert participants took part in our study; 2 astronauts, 5 engineers, 8 instructors, 2 managers and 3 scientists.}
\end{table*}

\subsection{Participants}
Given the niche character of our study, we approached potential participants by hand-picking and extending personal invitations to relevant experts in the human spaceflight field. 20 participants were recruited in total (see Table \ref{tab:participants} for details). 2 astronauts were included in our study, both of them currently active, having so far spent a combined duration of over 700 days in space. Both have also conducted a number of EVA operations outside the International Space Station (ISS) totaling over 39 hours. The instructors in our study have all experience with delivering astronaut training and performing operations planning. Several of them are likewise actively monitoring astronaut mission activities via the European Spacecraft Communicator (EUROCOM); an international flight control center responsible for direct communication with the ISS crew. The engineers have all experience providing technical support for the ISS, including overseeing the maintenance of several of its modules. The scientists have all firsthand experience with taking part in analogue field studies simulating lunar and martian missions. The managers are currently involved in relevant space system development projects in an administrative capacity. All of them have however also prior hands-on experience. Notably, manager 1 is an expert in astronaut EVA training and has experience using different types of spacesuits.


\subsection{Procedure}
Participants were invited individually to complete our VR scenario. Every session began with the given participant being briefed about the purpose of our study and the task to be carried out in VR, after which the participant was asked to fill-in a questionnaire on demographics and previous VR experience. This was followed by a quick demonstration of the VR controllers. Participants were then given freedom to navigate through the virtual environment and approach the container transportation mission in whichever way they preferred, as long as they did not stray too far away from the EL3 landing site. Drawing on the think aloud protocol \cite{ericsson1980verbal}, we encouraged our participants to verbalize their reasoning while completing the task. Once the payload containers were successfully retrieved and transported, we asked our participants to stay in the virtual environment for a little longer while answering a set of semi-structured interview questions. These questions were all open-ended and prompted participants to share their reflections and comment on their experience. In particular we enquired about various features of the lunar lander, such as the cargo unloading mechanism, antenna placement, the design of the cargo containers, the ladder and the transportation cart. Participants were also asked to identify potential safety hazards and suggest design improvements. Any unusual behaviors or actions exhibited by the participants were likewise explored. No time limits were applied during the study. Instead, we sought to provide each participant with as much (or as little) time as they needed to complete the scenario and to answer our questions. Consequently, the total length of the conducted sessions ranged widely from 40 to 80 minutes.

Participant responses were recorded in the form of notes, audio recordings and questionnaire replies. The data-set was then independently coded by three of our researchers, with any inconsistencies being addressed through a discussion and subsequently synthesized into a qualitative thematic analysis \cite{braun2006using}. We placed a particular focus on relevant design reflections brought up by our participants and on the specific aspects of our VR simulation that had helped elicit these.

\section{Findings}
All of our participants had prior experience with VR. Only three of them described their experience as either \textit{very limited} or \textit{limited} (Engineer 3, Instructor 4 and Instructor 8), while the rest considered their experience to be of a \textit{moderate} to \textit{very high} level (self-reported on a 5-point scale). None of the participants reported discomfort in the form of dizziness or motion sickness at any point during the study. Likewise, none of the participants had any significant difficulty coming to grips with the VR controllers and completing the scenario. 

All in all, the VR scenario was seen as highly accessible, being frequently described as “\textit{straightforward}” (Engineer 5), “\textit{quite realistic}” (Instructor 3) or “\textit{fun and immersive}” (Astronaut 1). As detailed in the remainder of this section, this, however, did not prevent our participants from making numerous critical reflections and bringing up potential design issues relevant to the depicted lunar surface solutions.

\subsection{Lights and Shadows} 
The unique lighting conditions on the Moon’s south pole, and their impact on future crew operations, formed perhaps the most persistently recurring theme in our participants’ accounts. The sun’s position near the lunar horizon resulted in a situation where every surface object and terrain feature cast protracted shadows, shrouding much of the environment in pitch darkness. As our participants made clear, this rendered even the most basic tasks, such as locomotion, noticeably more challenging. Scientist 2, for instance, argued the lack of visual cues in dark areas made it harder to infer how fast he was walking. Similarly, Engineer 5 pointed out that his own shadow in VR was deep enough to occlude any terrain inside it, which introduced the risk of tripping over boulders or other obstacles. This became a particularly pressing problem when participants walked in a direction away from the sun and thus cast a shadow in their own direction of movement. By eliciting such considerations around the impacts of severe shadowing, it was clear that the VR environment had succeeded in generating a non-intuitive cognitive load arising from the contrast to normal terrestrial experience of shadows.

Whilst our participants generally responded to this by exercising heightened caution while walking, Manager 1 went a step further. Instead of approaching the EL3 landing site along a straight path, he took a roundabout route, constantly scanning the terrain for illuminated patches, seeking to minimize his exposure to shadowed areas. When prompted to elaborate the reasons behind his approach, he referred to the limited intensity of his helmet light as a prime factor: \\

\textbf{Manager 1:} \textit{"As you can see, there are these rocks for example… if you bump into this, you could fall down or you could damage your suit. So what I tried to do was to go around in the light. I don't expect the astronauts to work in complete shadow. Normally, what they would do… during EVA’s on the space station they have lights on their helmets. This light will need to be more powerful than what I have here." }  \\

Although the built-in helmet light of the xEMU suit was generally seen as helpful, most participants shared Manager 1’s view that it would indeed need to be improved in order to adequately support future crews during their lunar extravehicular activities. It soon became clear that this was not simply an issue of limited light intensity, but likewise of a narrow beam angle. As noted by Instructor 5, while attempting to work in the lunar shadow, his entire field of view would shrink into the size of the light cone emitted by his helmet light. Instructor 2 echoed the same concern, arguing that a viable helmet light would thus need to be “more scattered” in order to cover as wide an area as possible.

Several other design concepts featured in our scenario were likewise brought up with regards to the need for better illumination. Scientist 3 and Instructor 4, for instance, argued that the EL3 landing legs could easily turn into a tripping hazard for anyone working in their immediate vicinity. Similarly, Engineer 4 noted that poor lighting conditions contribute to payload containers and other vital equipment being difficult to find if misplaced. The same concern was voiced by Engineer 2, who argued all mobile hardware, in particular the transportation cart, would need to be equipped with an artificial light source. The addition of LED bars, or other forms of light strips, was frequently brought up as a potential solution to such problems. Astronaut 1, meanwhile, cautioned against excessive use of such artificial lighting, arguing that this would lead to greater energy consumption and higher maintenance requirements. Instead, he proposed, astronauts would be better off with a handful of portable floodlights which could be carried around and positioned manually to counter any lighting situation. Instructor 7 suggested placing reflective tapes on strategic locations, such as the EL3 ladder, as another energy efficient alternative.

Whilst such solutions might help alleviate the pitch black lunar shadows, the blinding sunlight, aggravated by the Moon's atmosphere-less environment, proved an even more difficult hurdle to overcome. With the sun near the horizon, at eye level, our participants frequently found themselves staring straight into it, prompting numerous comments on the risk of being blinded and suffering momentary vision impairment. Drawing on his experience from real EVA operations, Astronaut 2 likened the situation to sitting in a car with a dirty windscreen when the headlights from a passing car suddenly appear. As he explained, this would not only blind the driver, but it would likewise make the windscreen “lit up and foggy”.

The sunlight reflecting from the aluminum and metallic components on the EL3 was also found problematic by many participants. Pointing to existing spacecraft solutions, Engineer 2 argued the use of anti-glare coating might be a potential remedy.   \\

\textbf{Engineer 2:} \textit{"If they have an option of covering some parts with a matte material, that would be nice. White textile, like we had on the ATV [Automated Transfer Vehicle], which is good for thermal control. Or you can also have a black coating, which is less reflective, like on the Soyuz."}   \\

Other participants advocated equipping the astronaut helmets with strong sun visors, however it was not immediately clear how practical such a solution would be given the frequent transitions between shadowed and illuminated areas. Just like Astronaut 2 before him, Instructor 5 employed a car-driving analogy to better explain this problem. When driving on a bright sunny day, he argued, the driver might benefit from wearing a pair of sunglasses. But upon entering a dark tunnel, the driver would be forced to take the sunglasses off and spend a few precious moments adapting to the new light setting.

Our VR scenario produced a number of instances where such sudden transitions became a matter of concern. For example, the near zero angle of sunlight resulted in half of the EL3 being occluded. When attempting to climb up on the \textit{dark} side of the lander, multiple participants reported that as soon as they reached the top of the ladder and gazed across the cargo deck, the sun would hit their face. The transition from pitch darkness to blinding light would in this sense be so rapid, and come at such a sensitive moment, that our participants saw it as imperative for future design efforts to take this into consideration.

The stark contrast between highlights and shadows conjured up by our VR simulation also proved a useful tool for visualizing temperature differences. Given their expertise, all of our participants were aware of the extreme thermal fluctuations on the lunar surface, with objects experiencing temperatures upwards 120 degrees Celsius when illuminated and below -200 degrees Celsius when occluded. Engineer 3 took this opportunity to reflect on the ideal placement for radiators and cooling systems on the EL3 landing vehicle, hypothesizing that a liquid circulatory system might prove effective in balancing the temperature differences across its body. Similarly, Scientist 1 cautioned against spending prolonged time in the shade, arguing the cold temperature would put unreasonable strain on sensitive cargo, such as biological materials. To tackle this issue, Engineer 2 proposed connecting cargo containers to an external power source via an \textit{umbilical cord}, allowing each container to be equipped with an active temperature regulation mechanism.

A closely related topic that also attracted multiple reflections was the issue of solar power generation. Upon examining the illuminated surfaces in VR, our participants frequently speculated about the ideal placement and angle of solar panels on the lander. Manager 1, for instance, suggested a situation might arise where the EL3 inadvertently lands in a shadow. To deal with such contingencies, he argued, it should be made possible to mount a solar panel on a transportation cart, enabling crews to move it into the sunlight to recharge mission critical equipment.   

\subsection{Dimensions and Ergonomics}
But it was not just the simulated light behavior that drew the attention of our participants. Having the option to experience prototypes interactively using an immersive 3D technology sparked numerous reflections concerning their dimensions and ergonomic appropriateness. This became perhaps most apparent during interactions with the cargo containers. Whilst most of our participants argued they would take advantage of a transportation cart if having to move the containers across larger distances (such as from the EL3 landing site to a habitation module), they also agreed that manual transportation across shorter distances (such as from the cart to an airlock chamber) would occasionally be unavoidable. Yet, upon spending some time manipulating and examining the containers in VR, the majority of our participants expressed skepticism regarding their viability in a real-world situation. Chiefly, the container handles did not appear to have been designed with EVA operations in mind. As Astronaut 2 noted, the thick spacesuit gloves worn during EVA’s are notorious for making otherwise trivial tasks \textit{“really tricky”}. Lifting up the cargo container using its relatively small handles was brought up as the most immediate example of this. Astronaut 1 echoed the same concerns, arguing all EVA interfaces would need to be somewhat overdimensioned in order to be compatible with astronaut activities:    \\

\textbf{Astronaut 1:} \textit{"Everything we do with our hands in spacesuits needs to be big. The gloves are bulky and have very little dexterity. Little to none. It’s actually one of the main concerns of any extravehicular activity, how dexterous you can be. So if you want to help and make a design that is a little more conducive to productivity and speed…  in this case I’d like to have a handle. A big fat handle that I could use while doing EVA. When I look at these, these are not EVA interfaces."} \\

Whilst our VR scenario was purely visual in nature, this did not prevent participants from attempting to infer the mass of objects, nor to speculate regarding the implications mass could have on an object’s manipulability. A large number of participants, for example, pointed out that carrying the weight of a container in one hand would shift astronauts’ center of gravity, potentially throwing them off balance. Manager 1 developed an alternative approach, grabbing two containers at the same time, one in each hand, arguing this would leave him more stable. He did admit, however, that two containers might prove too heavy to carry in spite of the reduced lunar gravity. Others frequently attempted to carry a single container in front of their bodies using both of their hands. Although this approach did seem to address their concerns regarding balance impairments, it also introduced another problem. As Engineer 1 discovered for himself, attempting to carry a container in this manner ended up partially obstructing his view:   \\

\textbf{Engineer 1:} \textit{"When I grab it using these side handles… A natural movement of my hands is like this [holds up the container vertically in front of his body], then I have it in front of me… see? But now it’s blocking a bit of my view."}   \\

Both Scientist 2 and Instructor 2 explicitly stated that maintaining clear visibility while carrying equipment ought to be a top design priority, citing loss of situational awareness and increased risk of tripping as potential consequences, should this requirement be breached. The majority of participants arrived at the same realization, prompting a search for possible workarounds. Once again, redesigning the container handles was brought up as the most readily available solution. As Astronaut 2 explained, when it comes to portable equipment, ideally “every surface should have a handle”, allowing astronauts to grab it from all sides. Increasing the number of handles would in this sense contribute to increased ergonomic flexibility by providing the user with more options.

Several alternative solutions were likewise brought up. Scientist 1, for instance, proposed an attachment mechanism that would allow astronauts to hook containers directly to their suits. Instructor 5 and Engineer 3, on the other hand, suggested putting wheels on each container, arguing that, much like a spinner luggage, dragging cargo containers would be easier than carrying them.

The shape of the cargo containers also attracted some criticism. Instructor 6 found that their cylindrical shape made containers prone to roll around, rather than sitting still once put down. Engineer 2 followed a similar line of reasoning. Upon unsuccessfully attempting to stack multiple containers on top of the cart, he suggested a rectangular shape would make containers more tileable, which might be useful for storage and transportation purposes.

The notion of having to use a ladder to climb up on top of the EL3 emerged as another major point of reflection, with participants typically asserting that additional safety measures would need to be implemented to prevent astronauts from falling. Snap hooks and carabiners were frequently argued to be of critical importance. Curiously, the astronauts partaking in our study did not share the same concerns, expressing greater worry for breaking vital equipment on the lander, rather than for their own safety. Astronaut 1, for instance, warned against the risk of inadvertently damaging a solar panel while climbing up, suggesting the gap between the ladder and the EL3 body is currently too small:   \\

\textbf{Astronaut 1:} \textit{"Right now I’m looking at the ladder and it looks almost 90 degrees vertical. Which is probably best for the launch configuration, but I could certainly see a very simple design where you pull the ladder into a 60 degree angle, or something like that, just to pull it away from the solar panels. My experience is that the pressurized spacesuits that we have for EVA’s… it’s really hard to see your feet. [starts climbing up the ladder] So there is a high chance here that my foot… my big booted foot can go through and hit that solar panel. So the ladder needs to be further away to make sure I don't hit it."}   \\

The concern that EVA suits might prevent astronauts from seeing their feet while climbing was also repeated by Engineer 2. To address this issue, he proposed equipping the top of the ladder with a set of side-view mirrors, much like the side-view mirrors that can be found on cars.

Operating on top of the lander was likewise seen as problematic. The cramped cargo deck offered little maneuvering room and, as noted by Instructor 5, there is a high risk one might inadvertently push over vital components when turning around while wearing a bulky spacesuit. Scientist 2 suggested a protective plexiglass dome should therefore be placed over the top of sensitive instruments, such as antennas. There was some disagreement whether additional steps should be taken to protect the astronauts while on top of the EL3. While Instructor 7 saw the need to implement a safety railing, Scientist 2 felt the existing cargo cradles would provide enough support for astronauts to hold on to.    \\

\subsection{The Bigger Picture}
As the study sessions progressed, many of our participants began shifting their focus from the individual design solutions featured in the scenario, to instead reflect on the role these solutions might play in the broader context of a lunar surface mission. By pondering potential synergies and points of friction, Engineer 2, for instance, suggested the EL3, with its robust solar panels, could be turned into a charging station, generating power for rovers and other key assets. Similarly, Scientist 1 proposed the lander might act as a provisional “cloud data center” collecting, storing and communicating mission relevant information to help coordinate nearby crews.

The extent and manner in which astronauts ought to get involved in surface operations was another frequently recurring theme in this vein of reflection. One aspect of our scenario that typically triggered such considerations was the cargo unloading mechanism, with our participants frequently contemplating who should be in charge of initiating the unloading sequence. Manager 1, for example, argued that it is the astronauts on the lunar surface who should be in control of such mechanisms due to their superior situational awareness. Engineer 3 concurred, arguing this would help prevent accidents, such as a cargo container being dropped on a bystanding astronaut.

Scientist 2, on the other hand, felt confident that such tasks could be safely delegated to the mission control center on Earth. This, he explained, would help lessen the astronaut’s workload. Engineer 2 adopted a similar stance, arguing that astronaut EVA-duration is constrained by a number of factors (such as limited oxygen supply) and initiating the cargo unloading sequence on the astronaut’s behalf would thus help save some of their precious time. Astronaut 2 appeared to share this position, explaining that the pre-planned nature of EVA operations typically leaves little room for astronaut’s own initiative. Nevertheless, he added, a backup option should always be in place in case of unforeseen circumstances:   \\

\textbf{Astronaut 2:} \textit{"We work with the ground [control center] all the time. So I think it's a good thing if the ground has control over it. But there should be maybe a plan B. It could be a contingency scenario that if something breaks, if we have to be autonomous on the Moon, then of course we need to be able to operate the important mechanisms. But I don't think that will ever be used. The ground will be running this operation. It's not like the astronauts say “okay, we're just gonna take this cargo off from the EL3 and put it over in the lab”. Instead it will be a scheduled activity with the whole control center sitting there on their toes, making sure that nothing breaks and everything works. So it will be a meticulous timeline, just like in an EVA on ISS."}   \\

Similarly, some of our participants, like Instructor 4, argued that all hazardous or physically demanding tasks ought to be offloaded to robotic solutions. Instructor 5 agreed, stating that any task that would not require “very precise action” should preferably be automated.

Such views, however, were far from uncontested. Instructor 7 and Engineer 1, for instance, stressed that the awareness and critical thinking possessed by human astronauts would introduce a degree of flexibility, making it easier to adjust procedures in response to a dynamic situation. Scientist 1 stated that robotic solutions are typically pre-programmed to carry out a narrowly defined task, lacking the versatility of humans. The breadth of tasks that could be carried out by trained human astronauts, he elaborated, makes them far more economical than relying on robotic solutions alone. Instructor 8 adopted a more measured stance, arguing that astronauts ought to be kept in the loop regarding all aspects of a surface operation, but their primary role should be one of oversight, rather than direct intervention. Assuming manual control over processes, she argued, should only be done when there is no other option.

All in all, VR proved useful in stimulating reflection on the optimal distribution of responsibilities and other factors instrumental for establishing an efficient collaborative ecosystem of humans and robotic systems on the Moon. 

\subsection{The Limits of VR}
While our scenario did succeed in eliciting a range of design considerations relevant to the EL3 and future lunar surface operations, it is crucial to acknowledge that several technical limitations likely also had a part in shaping the comments made by our participants.

The lack of haptic feedback emerged as arguably the most significant constraint, as became evident when lifting and manipulating cargo containers and other equipment. Nearly all participants noted the absence of weight, which, as Astronaut 2 and Engineer 4 elaborated, made it impossible to correctly infer the center of mass, inertia and rotational momentum of objects. Instructor 2, for example, argued that this had precluded him from providing detailed assessment of certain ergonomic factors, such as the ideal positioning of his arms while carrying containers.

A closely related problem was the inability of our VR simulation to recreate the movement constraints imposed by wearing a bulky xEMU suit. Scientist 3, for instance, was adamant about the unrealistic ease of player movement, while Instructor 4 and Scientist 2 both argued that properly evaluating some procedures, such as climbing a ladder, can only be done once the suit limitations are fully taken into account.

As Instructor 5 summarized, audiovisual VR simulations are therefore poorly equipped to facilitate assessments of questions concerning object manipulation and physical mechanics. Astronaut 2 reiterated the same argument, suggesting that the main strength of VR is rather to be found in its capacity to provide users with situational awareness:  \\

\textbf{Astronaut 2:} \textit{"It doesn't really tell me much about the mechanics of things. I mean, because of the limited haptic feedback. But situational awareness is something else. It’s good at giving the astronauts the awareness of what situation they would get themselves into… with the lighting, you know… I mean, I already, through that [the VR simulation], have a good feeling of what it would be like there [on the Moon]. That's a tremendous advantage. So I don't want to say it's not worth anything, it's worth a lot."}  \\

Similar views were echoed by other participants. Instructor 5 commended the opportunity to experience design solutions as components of a broader context, rather than as isolated prototypes. Instructor 4 felt this made VR particularly useful for planning of EVA operations. Engineer 3 concurred, finding VR useful to visualize timing of robotic actions and identify areas where human intervention might be needed.

Overall, it became clear that some of the limitations, at least in part ingrained in contemporary VR technology, could potentially undermine the viability of VR as a design tool. Equally important though, several strengths were identified that might prove highly useful during forthcoming design activities. In the following section we discuss these findings in more detail.

\section{Discussion}
\textcolor{black}{Interactive simulations in VR provide the means for rapid assessment and iterative development of novel design concepts without incurring many of the financial, logistical or temporal costs typically associated with real-world prototype deployments. In this study we set out to explore VR-enabled design processes in the unique context of the ongoing lunar landing programme. Using the EL3 project as our arena, we have carried out a simulated field study featuring an expert group of participants. }

Traditionally, approaches in this domain have relied on field studies at real-world analogue sites. By recreating some of the Moon's extreme environmental conditions, such studies have proven useful in providing an opportunity to evaluate aspects of the operational performance of a design solution under realistic circumstances. In addition, classical field studies have likewise been successfully used to assess the extent to which design solutions fit into a broader mission architecture. 

On the other hand, the high complexity and resource-intensive nature of such studies have also contributed to a situation where the number and frequency of experimental deployments tends to be limited and typically only involve a small number of participants, leading to slow feedback loops and rigid project structures. 

Our findings suggest that VR-based field studies could help mitigate some of these issues by enabling more agile user-centric design approaches commonly found in HCI. Similar to real-world analogue environments, VR is well suited for recreating certain aspects of the lunar environmental conditions, while also allowing the assessment of design concepts in a broader mission context. More than just eliciting actionable reflections, VR appears well qualified to facilitate team coordination and thus improve the overall project management processes in space systems design. Below we elaborate these key takeaways further and offer directions for future research.

\subsection{VR Enables User-Centered Design}

Our participant’s qualitative experiences have demonstrated the capacity of VR to elicit and invoke a range of reflections concerning ergonomics, safety and crew performance issues surrounding a low TRL concept. Although UCD approaches have already been employed with great success in other domains (e.g. automotive \cite{goedicke2018vr} or medical \cite{reinschluessel2017virtual} industries), \textcolor{black}{the reliance on analogue field studies in projects concerned with design of lunar surface solutions have resulted in user studies taking place late in the development cycle, at low frequencies and with a restricted number of participants.  }

\textcolor{black}{A strong argument for why VR might be capable of upending this established order is the near absolute control over light behavior offered by modern VR engines.} Given that the unique lunar lighting conditions will inevitably interface with nearly every facet of future human operations on the Moon, taking these into account during early stage prototyping and design activities will be critical. By granting users an opportunity to assess design concepts in any lighting situation immersively and interactively, VR thus appears uniquely qualified for this particular purpose. In doing so, VR-based field studies might open up a novel and highly resource efficient approach to collecting potentially vital user feedback from the very earliest stages of a lunar system's development. 

Equally important, however, is to take into account some of the limitations impairing the use of VR in other areas of design assessment. Accurate simulation of physical interactions in VR, for instance, remains a major technological challenge \cite{he2017physhare}. The lack of perceived lunar gravity, mass and inertia, along with spacesuit movement constraints, were explicitly identified as drawbacks by our participants, with some suggesting this might have resulted in potentially important design challenges being left underexposed in our study. \textcolor{black}{Whilst section \hyperref[sec:limitations]{6.4} elaborates potential ways of addressing some of these limitations, it seems likely the gap in fidelity between real and simulated field studies will never disappear entirely.  }

It is not our intention, then, to suggest VR-based design enquiries ought to replace classical analogue studies, \textcolor{black}{but rather to substitute for them during the early design stages when deployment} and testing of physical prototypes is not yet feasible. Our study has surfaced ample evidence suggesting this approach might help produce prototypes for analogue deployments that are more refined and conducive with user needs, thus saving time and resources in the overall development process.  

\textcolor{black}{Going back to Trotta et al.’s call for the HCI community to innovate the design of future space interfaces \cite{trotta2020communicating}, we can thus conclude that VR has the potential to significantly broaden the scope of applicability of HCI design methods, making user-centered design processes viable in the realm of space systems design. }

\subsection{VR Facilitates Contextual Inquiry}
\textcolor{black}{Apart from being suitable for interactive visualization of prospective design solutions in the context of authentic environmental conditions (e.g. lighting), our study likewise found VR useful for facilitating reflections concerning the broader operational context, including the interoperability between different design concepts and their fit with potential mission procedures.  }

The need to contextualize early-stage design prototypes has drawn some attention in HCI already. Salovaara et al., for instance, described a “blind spot” in HCI research, stemming from the fact that any future-oriented prototype deployed in a study still exists firmly in the present world \cite{Salovaara2017}. Similarly, Lindley et al. have advocated the need to evaluate early design concepts as they might exist once fully adopted, “beyond their prototypical implementation” \cite{Lindley2017}, in order to better anticipate (and evade) potential frictions surrounding their future use.   

\textcolor{black}{One popular way of tackling this issue has is the use narrative prototypes, such as scenarios or design fictions. For, as Bleecker elaborated, \textit{“a story about future technology is also a story about the broader social practices we imagine growing up around it.”} \cite{bleecker2022design} }

\textcolor{black}{Such a ‘big picture’ lens is also traditionally of great interest during analogue field studies and experimental deployments of prospective space solutions. As Beaton et al. explain, a key goal of such studies is to identify and assess design elements in the context of \textit{“the organization and flow of personnel, communications, hardware, software, and data products.”} \cite{beaton2019b}  }

\textcolor{black}{Correspondingly, by instantiating design scenarios in VR, we were able to simultaneously convey immersive representations of concepts as well as contexts, making them available for user exploration and examination. In doing so, VR effectively endowed our design assessments with elements of contextual inquiry \cite{duda2020contextual}. The benefits of this quickly became apparent, as our study surfaced numerous instances of synergies and frictions arising between the depicted design solutions. Examples include compatibility issues between portable cargo containers and EVA suits, the sharing of information between astronauts and the lunar lander module, or the optimal distribution of competencies between mission control center and lunar surface crews. It is improbable that either of these considerations would have emerged by studying a prototype alone, without simulating a broader operational context. Yet, each of these considerations could prove critical for informing and steering future design efforts. }

\textcolor{black}{The value of such contextual inquiries extends beyond the space systems design domain. By way of example, previous research noted there is a lack of tools and methods for evaluating multi-device experiences \cite{dong2016understanding}. One of the contributions of this work, then, is that it provides an empirical rationale for VR-based scenarios in design assessments of complex and context-dependent multi-device human-machine ecosystems.}

\subsection{VR Coordinates Project Teams} 
As previously mentioned, space systems design projects are notoriously rigid, having to deal with slow feedback loops that contribute to poor coordination with end users (e.g. astronauts) and other stakeholders. Typically, efforts to tackle such issues have crystallized in various forms of Model Based Systems Engineering (MBSE). The core idea behind MBSE is to center engineering projects around a set of models that capture key project aspects by highlighting important subsystems and elements while simplifying or omitting less relevant features \cite{jacobson1999}. Such models can feature both social and technological contexts, helping engineers to coordinate and incorporate an extensive and wide-ranging set of requirements, including flexibility, sustainability, real-time capability, adaptability, expandability, reliability or usability \cite{rhodes2008addressing}.

By replacing inflexible, costly, and slow paper documentation with digital communication pivoting on digital twins and digital threads, MBSE facilitates rapid information exchange and collaboration between several individuals or teams, allowing for design analysis activities from the very beginning of an engineering project.

The adoption of MBSE-centric approaches has in this sense been hailed as a potential shift away from the classical, highly developed, linear, rigid and document-centric processes, towards a more collaborative, non-linear and agile digital information-based approach.

Nevertheless, contemporary MBSE approaches and the associated best practices vary greatly across different engineering disciplines and the ideal approach by which a model should be conveyed remains a frequent topic of debate \cite{ESA_MBSE2021}. Consequently, interdisciplinary teams are frequently facing a situation where the time and effort required by training, along with the need for a deep understanding of the MBSE modeling language, the complexity of the models adopted and the limited flexibility of available tools makes its use difficult. As a result, the MBSE design approach has yet to see any widespread adoption \cite{Lindblad}, \textcolor{black}{particularly so in human spaceflight-related projects notorious for having to balance engineering, scientific and political interests \cite{ESA_MBSE2021}.   }

We would suggest that one of the contributions of our work is that it demonstrates potential applicability of VR as a viable vehicle for MBSE approaches in this domain. \textcolor{black}{System models conceived via an MBSE approach can indeed be virtualized using VR tools from the earliest design stages, making desired project outcomes available for review by relevant stakeholders in a concurrent manner. This allows to lower the efforts and stress levels of the development team while also reducing the risk of cost overruns and project delays.  }

Furthermore, our work indicates that VR might likewise help address the lamented lack of MBSE interoperability, which has been identified as a key requirement to fully adopt a model-based approach in the context of human spaceflight projects \cite{Lindblad, Eisenmann}. As shown by our findings, early conceptual models conveyed by means of VR scenarios can realistically be employed to engage a diverse group of stakeholders, ranging from astronauts to engineers and managers, establishing a unifying language that makes early-stage design solutions, their co-creation and assessments, accessible across fields and disciplines.

\subsection{Limitations and Future Directions} 
\label{sec:limitations}
\textcolor{black}{Unlike VR-based training applications typically striving to maximize user’s \textit{behavioral} validity \cite{shaw2019heat}, the prime objective of our simulated field study was rather to elicit valid user \textit{reflections}. A case could then be made that the phenomenological nature of our enquiry, with expert users drawing on their own personal experience, has permitted some level of abstraction and simplification with regards to the simulated interactions. Past studies seem to support this; Osterlund et al., for instance, found that a purely audiovisual simulation of aerospace maintenance operations in protective suits could elicit valid and actionable design feedback from subject matter experts \cite{osterlund2011}.  }

\textcolor{black}{Whilst it is therefore unlikely that the monosensory nature of our VR simulation impeded the overall validity of the presented findings, we did also come across several instances where a broader sensory engagement delivered via a mixed reality interface would have likely resulted in richer and more accurate user feedback. Below, we discuss key problem areas in this vein and offer our recommendations concerning the development of future interfaces for simulation of lunar surface scenarios.}

  \begin{itemize}
\item {\textcolor{black}{\textbf{Haptics} - The lack of physical properties, including a sense of touch, weight and rigidity hampered our participant’s ability to assess key ergonomic aspects surrounding the manipulation of cargo containers and other artifacts in our simulation. Future studies should thus consider augmenting key virtual objects of interest with real-world physical mockups using VR or Mixed Reality tools. This could, for example, be done by utilizing optical flow to track and superimpose the virtual environment on the mockups. Previous work has demonstrated that employing such mockups in conjunction with visual VR interfaces in this manner greatly increases user’s perceived immersion \cite{fitzgerald2018mediate, dangxiao2019haptic}. }} \\

\item {\textcolor{black}{\textbf{Reduced gravity} - Future work should also explore possible coupling of VR simulations with gravity offload systems to artificially reduce gravity levels - both that of the user and the aforementioned physical mockups. Such systems are generally characterized by anchor points onto which physical mockups can be attached in order to reduce their weight by a desirable factor (see NASA ARGOS system for an example \cite{orr2022effects}). Similarly, applying such systems to the users would allow them to better experience the sensation of hypogravity when walking around.  }}\\

\item {\textcolor{black}{\textbf{Spacesuits} - Accurate simulation of movement constraints imposed by EVA suits would likewise help enhance the validity of relevant design reflections. Whilst a range of movement constraining solutions has already been employed in real-world training situations \cite{braden2002development}, their possible use with VR remains largely unexplored, likely due to limited applicability and tracking difficulties. An interim solution might be to utilize real EVA suits in Cave Automatic Virtual Environments (CAVEs). We would suggest, however, that this approach is not ideal, given the limited immersive capabilities of CAVE’s with respect to more modern VR headsets. Instead, we would advocate the integration of VR goggles into an adapted EVA suit, thus achieving both accurate movement limitations and visual immersion. With such an interface, it is likely that the bulky nature of EVA gloves would impede the accuracy of optical finger tracking (even with visual tracking markers being employed). The simulation of scenarios involving fine motor skills would thus likely require gloves with built-in inertial tracking sensors. }}\\

\item {\textcolor{black}{\textbf{Communication} - The important role of communication between mission control center and lunar ground crews was frequently brought up by our study participants. Future studies exploring more complex lunar scenarios (e.g. work procedures dependent on active guidance from the mission control center) might be enhanced through interfaces capable of mimicking some of the relevant operational constraints, such as communication delays (typically cca 2,5 seconds for the Moon), to assess the impact this might have on crew workflows and explore potential remedies. } }\\

\item {\textcolor{black}{\textbf{Fused Reality} - Ultimately, drawing on all the points above, studies prioritizing realism might seek to reap the benefits of combining mixed reality interfaces with classical field studies. Virtual elements could then be used to enhance the analogue environment’s capacity to approximate lunar conditions, such as by simulating poor visibility, dust, or unique lunar terrain features. The notion of fusing real and virtual environments has already shown promise in practice. Military pilots, for instance, have made use of VR headsets while flying real planes in order to train docking procedures with a computer-generated tanker during mid-flight refueling \cite{conner2015fused}. } }
\end{itemize}

 \textcolor{black}{
This is by no means an exhaustive list of relevant problem areas, but rather an overview of those that most prominently surfaced through our study. Other work has pointed out, for instance, the benefits of multi user VR interfaces \cite{kopec2021participatory, berni2020applications}. Nevertheless, it is apparent that tackling either of the issues above would improve the overall simulation validity and thus help improve future VR-based enquiries into design challenges surrounding human lunar (and planetary) exploration.  }

\section{Conclusions}
\textcolor{black}{Our virtual field study has demonstrated the feasibility of employing a human-centered design approach during early stages of lunar system development. }The benefits of involving end-users and other stakeholders in such early design stages are numerous, ranging from cost and time savings to improved project management processes.   

We have argued that the flexibility of VR, coupled with its capacity to accurately simulate lunar lighting conditions, constitute a powerful incentive for its adoption during early system design stages. By simultaneously conveying concept and context, and thus putting the spotlight on interoperability of design solutions, VR was likewise found well suited for assessing synergies and frictions between design concepts and key mission elements. Moreover, our findings indicate that the visual nature and the accessibility of VR could make it a potent tool in facilitating communication between relevant team members from the earliest stages of development, thus bringing agile elements into traditionally rigid project structures. 

Based on our observations and participant feedback, we have likewise formulated key technical limitations related to VR-based simulations of lunar environments. In response, we have proposed possible ways to address these challenges through greater utilization of mixed reality interfaces. 

\textcolor{black}{Finally, an important ambition of this work has been to explore the usefulness of HCI research and design methods in the context of human lunar exploration.} Whilst traditionally being associated with the decidedly more down-to-earth setting of everyday \textit{life, work and play} \cite{deterding2011gamification}, HCI is well qualified to tackle many of the challenges surrounding humanity’s return to the Moon. The scale and scope of this endeavour can only be faced by bridging disciplines and drawing on their compound power. With this paper we have aspired to take a first step towards this goal.

\begin{acks}
We would like to thank the European Astronaut Center in Cologne for providing us with expert participants and making our study possible. In particular we are grateful to Lionel Ferra and the XR Lab for supplying us with technical resources for our work, to Ludovic Duvet for overseeing the development of our virtual EL3 mockup and to Hervé Stevenin for sharing material and insights concerning ESA’s analogue studies. We would also like to extend our thanks to the anonymous CHI reviewers who helped us think more deeply about our work and its significance in the context of human-computer interaction. 
\end{acks}

\balance






\bibliographystyle{ACM-Reference-Format}
\bibliography{sample-base}


\begin{thebibliography}{95}


\ifx \showCODEN    \undefined \def \showCODEN     #1{\unskip}     \fi
\ifx \showDOI      \undefined \def \showDOI       #1{#1}\fi
\ifx \showISBNx    \undefined \def \showISBNx     #1{\unskip}     \fi
\ifx \showISBNxiii \undefined \def \showISBNxiii  #1{\unskip}     \fi
\ifx \showISSN     \undefined \def \showISSN      #1{\unskip}     \fi
\ifx \showLCCN     \undefined \def \showLCCN      #1{\unskip}     \fi
\ifx \shownote     \undefined \def \shownote      #1{#1}          \fi
\ifx \showarticletitle \undefined \def \showarticletitle #1{#1}   \fi
\ifx \showURL      \undefined \def \showURL       {\relax}        \fi
\providecommand\bibfield[2]{#2}
\providecommand\bibinfo[2]{#2}
\providecommand\natexlab[1]{#1}
\providecommand\showeprint[2][]{arXiv:#2}

\bibitem[via(2017)]%
        {via_satellite_2017}
 \bibinfo{year}{2017}\natexlab{}.
\newblock \bibinfo{title}{Lockheed Martin on cutting costs with virtual reality
  - via satellite -}.
\newblock
\newblock
\urldef\tempurl%
\url{https://www.satellitetoday.com/innovation/2017/04/20/lockheed-martin-cutting-costs-virtual-reality/}
\showURL{%
\tempurl}


\bibitem[Beaton et~al\mbox{.}(2019)]%
        {beaton2019b}
\bibfield{author}{\bibinfo{person}{Kara~H Beaton}, \bibinfo{person}{Steven~P
  Chappell}, \bibinfo{person}{Andrew~FJ Abercromby}, \bibinfo{person}{Matthew~J
  Miller}, \bibinfo{person}{Shannon~E Kobs~Nawotniak},
  \bibinfo{person}{Allyson~L Brady}, \bibinfo{person}{Adam~H Stevens},
  \bibinfo{person}{Samuel~J Payler}, \bibinfo{person}{Scott~S Hughes}, {and}
  \bibinfo{person}{Darlene~SS Lim}.} \bibinfo{year}{2019}\natexlab{}.
\newblock \showarticletitle{Using science-driven analog research to investigate
  extravehicular activity science operations concepts and capabilities for
  human planetary exploration}.
\newblock \bibinfo{journal}{\emph{Astrobiology}}  \bibinfo{volume}{33}
  (\bibinfo{year}{2019}), \bibinfo{pages}{300--320}.
\newblock
Issue 4.


\bibitem[Beaton et~al\mbox{.}(2020)]%
        {Beaton2020}
\bibfield{author}{\bibinfo{person}{Kara~H Beaton}, \bibinfo{person}{Steven~P
  Chappell}, \bibinfo{person}{Alex Menzies}, \bibinfo{person}{Victor Luo},
  \bibinfo{person}{So~Young Kim-Castet}, \bibinfo{person}{Dava Newman},
  \bibinfo{person}{Jeffrey Hoffman}, \bibinfo{person}{Johannes Norheim},
  \bibinfo{person}{Eswar Anandapadmanaban}, \bibinfo{person}{Stewart~P
  Abercrombie}, \bibinfo{person}{Shannon E~Kobs Nawotniak},
  \bibinfo{person}{Andrew F~J Abercromby}, {and} \bibinfo{person}{Darlene S~S
  Lim}.} \bibinfo{year}{2020}\natexlab{}.
\newblock \showarticletitle{Mission enhancing capabilities for science-driven
  exploration extravehicular activity derived from the NASA BASALT research
  program}.
\newblock \bibinfo{journal}{\emph{Planetary and Space Science}}
  \bibinfo{volume}{193} (\bibinfo{year}{2020}), \bibinfo{pages}{105003}.
\newblock
\showISSN{0032-0633}
\urldef\tempurl%
\url{https://doi.org/10.1016/j.pss.2020.105003}
\showDOI{\tempurl}


\bibitem[Berg and Vance(2017)]%
        {berg2017industry}
\bibfield{author}{\bibinfo{person}{Leif~P Berg} {and} \bibinfo{person}{Judy~M
  Vance}.} \bibinfo{year}{2017}\natexlab{}.
\newblock \showarticletitle{Industry use of virtual reality in product design
  and manufacturing: a survey}.
\newblock \bibinfo{journal}{\emph{Virtual reality}} \bibinfo{volume}{21},
  \bibinfo{number}{1} (\bibinfo{year}{2017}), \bibinfo{pages}{1--17}.
\newblock


\bibitem[Berni and Borgianni(2020)]%
        {berni2020applications}
\bibfield{author}{\bibinfo{person}{Aurora Berni} {and} \bibinfo{person}{Yuri
  Borgianni}.} \bibinfo{year}{2020}\natexlab{}.
\newblock \showarticletitle{Applications of virtual reality in engineering and
  product design: Why, what, how, when and where}.
\newblock \bibinfo{journal}{\emph{Electronics}} \bibinfo{volume}{9},
  \bibinfo{number}{7} (\bibinfo{year}{2020}), \bibinfo{pages}{1064}.
\newblock


\bibitem[Bessone et~al\mbox{.}(2015)]%
        {Bessone2015}
\bibfield{author}{\bibinfo{person}{L. Bessone}, \bibinfo{person}{F. Sauro},
  {and} \bibinfo{person}{H. Stevenin}.} \bibinfo{year}{2015}\natexlab{}.
\newblock \showarticletitle{Training Safe and Effective Spaceflight Operations
  Using Terrestrial Analogues}.
\newblock \bibinfo{journal}{\emph{Space Safety is No Accident}},
  \bibinfo{pages}{313--318}.
\newblock
\urldef\tempurl%
\url{https://doi.org/10.1007/978-3-319-15982-9_37}
\showDOI{\tempurl}


\bibitem[Bleecker(2022)]%
        {bleecker2022design}
\bibfield{author}{\bibinfo{person}{Julian Bleecker}.}
  \bibinfo{year}{2022}\natexlab{}.
\newblock \showarticletitle{Design fiction: A short essay on design, science,
  fact, and fiction}.
\newblock \bibinfo{journal}{\emph{Machine Learning and the City: Applications
  in Architecture and Urban Design}} (\bibinfo{year}{2022}),
  \bibinfo{pages}{561--578}.
\newblock


\bibitem[Boy and Platt(2013)]%
        {boy2013situation}
\bibfield{author}{\bibinfo{person}{Guy~Andre Boy} {and} \bibinfo{person}{Donald
  Platt}.} \bibinfo{year}{2013}\natexlab{}.
\newblock \showarticletitle{A situation awareness assistant for human deep
  space exploration}. In \bibinfo{booktitle}{\emph{International Conference on
  Human-Computer Interaction}}. Springer, \bibinfo{pages}{629--636}.
\newblock


\bibitem[Braden and Akin(2002)]%
        {braden2002development}
\bibfield{author}{\bibinfo{person}{Jeffrey~R Braden} {and}
  \bibinfo{person}{David~L Akin}.} \bibinfo{year}{2002}\natexlab{}.
\newblock \showarticletitle{Development and testing of a space suit analogue
  for neutral buoyancy eva research}.
\newblock \bibinfo{journal}{\emph{Development}}  \bibinfo{volume}{1}
  (\bibinfo{year}{2002}), \bibinfo{pages}{2364}.
\newblock


\bibitem[Braun and Clarke(2006)]%
        {braun2006using}
\bibfield{author}{\bibinfo{person}{Virginia Braun} {and}
  \bibinfo{person}{Victoria Clarke}.} \bibinfo{year}{2006}\natexlab{}.
\newblock \showarticletitle{Using thematic analysis in psychology}.
\newblock \bibinfo{journal}{\emph{Qualitative research in psychology}}
  \bibinfo{volume}{3}, \bibinfo{number}{2} (\bibinfo{year}{2006}),
  \bibinfo{pages}{77--101}.
\newblock


\bibitem[Budzyn et~al\mbox{.}(2018)]%
        {Dorota}
\bibfield{author}{\bibinfo{person}{Dorota Budzyn}, \bibinfo{person}{Hervé
  Stevenin}, \bibinfo{person}{Matthias Maurer}, \bibinfo{person}{Francesco
  Sauro}, {and} \bibinfo{person}{Loredana Bessone}.}
  \bibinfo{year}{2018}\natexlab{}.
\newblock \showarticletitle{Prototyping of Lunar surface geological sampling
  tools for Moon spacewalk simulations by ESA}.
\newblock \bibinfo{journal}{\emph{69th International Astronautical Congress
  (IAC), Bremen, Germany}}.
\newblock


\bibitem[Cardano et~al\mbox{.}(2009)]%
        {cardano2009vr}
\bibfield{author}{\bibinfo{person}{M Cardano}, \bibinfo{person}{M Ferrino},
  \bibinfo{person}{M Costa}, {and} \bibinfo{person}{P Giorgi}.}
  \bibinfo{year}{2009}\natexlab{}.
\newblock \showarticletitle{VR/AR tools to support on orbit crew operations and
  P/Ls maintenance in the ISS pressurized Columbus module}. In
  \bibinfo{booktitle}{\emph{60th International Astronautical Congress, Daejeon,
  Republic of Korea}}. \bibinfo{pages}{12--16}.
\newblock


\bibitem[Carey et~al\mbox{.}(2021)]%
        {Carey2021}
\bibfield{author}{\bibinfo{person}{William~C Carey}, \bibinfo{person}{Ludovic
  Duvet}, \bibinfo{person}{Nick Gollins}, \bibinfo{person}{Rogier Schonenborg},
  \bibinfo{person}{Alexander Cropp}, \bibinfo{person}{Giorgio Cifani},
  \bibinfo{person}{Keith Stephenson}, \bibinfo{person}{Philipp~B Hager},
  \bibinfo{person}{Yannick~Le Deuff}, \bibinfo{person}{Kim Nergaard},
  \bibinfo{person}{Thorsten Graber}, \bibinfo{person}{Jennifer Reynolds},
  \bibinfo{person}{Giorgio Magistrati}, \bibinfo{person}{Sandra Magunsong},
  \bibinfo{person}{Jorge Alves}, \bibinfo{person}{Francesca McDonald}, {and}
  \bibinfo{person}{Nadine Boersma}.} \bibinfo{year}{2021}\natexlab{}.
\newblock \showarticletitle{European access to the lunar surface: EL3 mission
  options}.
\newblock \bibinfo{journal}{\emph{72nd International Astronautical Congress
  (IAC)}}.
\newblock


\bibitem[Carroll(2003)]%
        {carroll2003making}
\bibfield{author}{\bibinfo{person}{John~M Carroll}.}
  \bibinfo{year}{2003}\natexlab{}.
\newblock \bibinfo{booktitle}{\emph{Making use: scenario-based design of
  human-computer interactions}}.
\newblock \bibinfo{publisher}{MIT press}.
\newblock


\bibitem[Casini et~al\mbox{.}(2020)]%
        {casini2020lunar}
\bibfield{author}{\bibinfo{person}{Andrea~EM Casini}, \bibinfo{person}{Petra
  Mittler}, \bibinfo{person}{Aidan Cowley}, \bibinfo{person}{Lukas
  Schl{\"u}ter}, \bibinfo{person}{Marthe Faber}, \bibinfo{person}{Beate
  Fischer}, \bibinfo{person}{Melanie von~der Wiesche}, {and}
  \bibinfo{person}{Matthias Maurer}.} \bibinfo{year}{2020}\natexlab{}.
\newblock \showarticletitle{Lunar analogue facilities development at EAC: the
  LUNA project}.
\newblock \bibinfo{journal}{\emph{Journal of Space Safety Engineering}}
  \bibinfo{volume}{7}, \bibinfo{number}{4} (\bibinfo{year}{2020}),
  \bibinfo{pages}{510--518}.
\newblock


\bibitem[Casini et~al\mbox{.}(2018)]%
        {casini2018analysis}
\bibfield{author}{\bibinfo{person}{Andrea E~M Casini}, \bibinfo{person}{Paolo
  Maggiore}, \bibinfo{person}{Nicole Viola}, \bibinfo{person}{Valter Basso},
  \bibinfo{person}{Marinella Ferrino}, \bibinfo{person}{Jeffrey~A Hoffman},
  {and} \bibinfo{person}{Aidan Cowley}.} \bibinfo{year}{2018}\natexlab{}.
\newblock \showarticletitle{Analysis of a Moon outpost for Mars enabling
  technologies through a Virtual Reality environment}.
\newblock \bibinfo{journal}{\emph{Acta Astronautica}}  \bibinfo{volume}{143}
  (\bibinfo{year}{2018}), \bibinfo{pages}{353--361}.
\newblock


\bibitem[Chavers(2016)]%
        {chavers2016nasa}
\bibfield{author}{\bibinfo{person}{Donald~G Chavers}.}
  \bibinfo{year}{2016}\natexlab{}.
\newblock \showarticletitle{NASA lander technologies project status}.
\newblock In \bibinfo{booktitle}{\emph{AIAA space 2016}}.
  \bibinfo{pages}{5221}.
\newblock


\bibitem[Clynes and Kline(1960)]%
        {clynes1960cyborgs}
\bibfield{author}{\bibinfo{person}{Manfred~E Clynes} {and}
  \bibinfo{person}{Nathan~S Kline}.} \bibinfo{year}{1960}\natexlab{}.
\newblock \showarticletitle{Cyborgs and space}.
\newblock \bibinfo{journal}{\emph{Astronautics}} \bibinfo{volume}{14},
  \bibinfo{number}{9} (\bibinfo{year}{1960}), \bibinfo{pages}{26--27}.
\newblock


\bibitem[Conner(2015)]%
        {conner2015fused}
\bibfield{author}{\bibinfo{person}{Monroe Conner}.}
  \bibinfo{year}{2015}\natexlab{}.
\newblock \bibinfo{title}{Fused Reality: Making the Imagined Seem Real}.
\newblock
\newblock


\bibitem[Dangxiao et~al\mbox{.}(2019)]%
        {dangxiao2019haptic}
\bibfield{author}{\bibinfo{person}{Wang Dangxiao}, \bibinfo{person}{GUO Yuan},
  \bibinfo{person}{LIU Shiyi}, \bibinfo{person}{Yuru Zhang},
  \bibinfo{person}{Xu Weiliang}, {and} \bibinfo{person}{Xiao Jing}.}
  \bibinfo{year}{2019}\natexlab{}.
\newblock \showarticletitle{Haptic display for virtual reality: progress and
  challenges}.
\newblock \bibinfo{journal}{\emph{Virtual Reality \& Intelligent Hardware}}
  \bibinfo{volume}{1}, \bibinfo{number}{2} (\bibinfo{year}{2019}),
  \bibinfo{pages}{136--162}.
\newblock


\bibitem[de~Haan(2015)]%
        {de2015hci}
\bibfield{author}{\bibinfo{person}{Geert de Haan}.}
  \bibinfo{year}{2015}\natexlab{}.
\newblock \showarticletitle{HCI Design Methods: where next? from user-centred
  to creative design and beyond}. In \bibinfo{booktitle}{\emph{Proceedings of
  the European Conference on Cognitive Ergonomics 2015}}.
  \bibinfo{pages}{1--8}.
\newblock


\bibitem[De~Monchaux(2011)]%
        {de2011spacesuit}
\bibfield{author}{\bibinfo{person}{Nicholas De~Monchaux}.}
  \bibinfo{year}{2011}\natexlab{}.
\newblock \bibinfo{booktitle}{\emph{Spacesuit: Fashioning Apollo}}.
\newblock \bibinfo{publisher}{MIT press}.
\newblock


\bibitem[Deterding et~al\mbox{.}(2011)]%
        {deterding2011gamification}
\bibfield{author}{\bibinfo{person}{Sebastian Deterding},
  \bibinfo{person}{Miguel Sicart}, \bibinfo{person}{Lennart Nacke},
  \bibinfo{person}{Kenton O'Hara}, {and} \bibinfo{person}{Dan Dixon}.}
  \bibinfo{year}{2011}\natexlab{}.
\newblock \showarticletitle{Gamification. using game-design elements in
  non-gaming contexts}.
\newblock In \bibinfo{booktitle}{\emph{CHI'11 extended abstracts on human
  factors in computing systems}}. \bibinfo{pages}{2425--2428}.
\newblock


\bibitem[Dong et~al\mbox{.}(2016)]%
        {dong2016understanding}
\bibfield{author}{\bibinfo{person}{Tao Dong}, \bibinfo{person}{Elizabeth~F
  Churchill}, {and} \bibinfo{person}{Jeffrey Nichols}.}
  \bibinfo{year}{2016}\natexlab{}.
\newblock \showarticletitle{Understanding the challenges of designing and
  developing multi-device experiences}. In
  \bibinfo{booktitle}{\emph{Proceedings of the 2016 ACM Conference on Designing
  Interactive Systems}}. \bibinfo{pages}{62--72}.
\newblock


\bibitem[Drey et~al\mbox{.}(2020)]%
        {drey2020vrsketchin}
\bibfield{author}{\bibinfo{person}{Tobias Drey}, \bibinfo{person}{Jan
  Gugenheimer}, \bibinfo{person}{Julian Karlbauer}, \bibinfo{person}{Maximilian
  Milo}, {and} \bibinfo{person}{Enrico Rukzio}.}
  \bibinfo{year}{2020}\natexlab{}.
\newblock \showarticletitle{Vrsketchin: Exploring the design space of pen and
  tablet interaction for 3d sketching in virtual reality}.
\newblock \bibinfo{journal}{\emph{Proceedings of the 2020 CHI Conference on
  Human Factors in Computing Systems}}, \bibinfo{pages}{1--14}.
\newblock


\bibitem[Duda et~al\mbox{.}(2020)]%
        {duda2020contextual}
\bibfield{author}{\bibinfo{person}{Sabrina Duda}, \bibinfo{person}{Carolyn
  Warburton}, {and} \bibinfo{person}{Nissa Black}.}
  \bibinfo{year}{2020}\natexlab{}.
\newblock \showarticletitle{Contextual research}. In
  \bibinfo{booktitle}{\emph{International Conference on Human-Computer
  Interaction}}. Springer, \bibinfo{pages}{33--49}.
\newblock


\bibitem[Duvet et~al\mbox{.}(2021)]%
        {Duvet}
\bibfield{author}{\bibinfo{person}{Ludovic Duvet}, \bibinfo{person}{Alexander
  Cropp}, \bibinfo{person}{Philipp Hager}, \bibinfo{person}{Kim Nergaard},
  \bibinfo{person}{Nick Gollins}, \bibinfo{person}{Giorgio Cifani},
  \bibinfo{person}{Rogier Schonenborg}, \bibinfo{person}{William Carey},
  \bibinfo{person}{Florian Renk}, \bibinfo{person}{Lorenzo Bucci},
  \bibinfo{person}{Sarah Mangunsong}, \bibinfo{person}{Yannick~Le Deuff},
  \bibinfo{person}{Alberto~Gonzalez Fernandez}, \bibinfo{person}{Antonios
  Tavoularis}, \bibinfo{person}{Andrea Casini}, \bibinfo{person}{Gilles
  Nzokira}, \bibinfo{person}{Giorgio Magistrati}, \bibinfo{person}{Davide
  Rovelli}, {and} \bibinfo{person}{Keith Stephenson}.}
  \bibinfo{year}{2021}\natexlab{}.
\newblock \showarticletitle{European Access to the Lunar Surface: EL3}.
\newblock \bibinfo{journal}{\emph{72nd International Astronautical Congress
  (IAC)}}.
\newblock


\bibitem[Eisenmann et~al\mbox{.}(2009)]%
        {Eisenmann}
\bibfield{author}{\bibinfo{person}{Harald Eisenmann}, \bibinfo{person}{Juan
  Miro}, {and} \bibinfo{person}{Hans~Peter de Koning}.}
  \bibinfo{year}{2009}\natexlab{}.
\newblock \showarticletitle{MBSE for European Space-Systems Development}.
\newblock \bibinfo{journal}{\emph{INSIGHT}}  \bibinfo{volume}{12}
  (\bibinfo{date}{12} \bibinfo{year}{2009}), \bibinfo{pages}{47--53}.
\newblock
Issue 4.
\showISSN{2156485X}
\urldef\tempurl%
\url{https://doi.org/10.1002/inst.200912447}
\showDOI{\tempurl}


\bibitem[Eppler(1991)]%
        {eppler1991lighting}
\bibfield{author}{\bibinfo{person}{Dean~B Eppler}.}
  \bibinfo{year}{1991}\natexlab{}.
\newblock \showarticletitle{Lighting constraints on lunar surface operations}.
\newblock \bibinfo{journal}{\emph{NASA STI/Recon Technical Report N}}
  \bibinfo{volume}{91} (\bibinfo{year}{1991}), \bibinfo{pages}{23014}.
\newblock


\bibitem[Ericsson and Simon(1980)]%
        {ericsson1980verbal}
\bibfield{author}{\bibinfo{person}{K~Anders Ericsson} {and}
  \bibinfo{person}{Herbert~A Simon}.} \bibinfo{year}{1980}\natexlab{}.
\newblock \showarticletitle{Verbal reports as data.}
\newblock \bibinfo{journal}{\emph{Psychological review}}  \bibinfo{volume}{87}
  (\bibinfo{year}{1980}), \bibinfo{pages}{215}.
\newblock
Issue 3.


\bibitem[ESA(2021a)]%
        {LESA}
\bibfield{author}{\bibinfo{person}{ESA}.} \bibinfo{year}{2021}\natexlab{a}.
\newblock \bibinfo{title}{As above, so below – lunar rescue device tested
  underwater}.
\newblock
\newblock
\urldef\tempurl%
\url{https://www.esa.int/Science_Exploration/Human_and_Robotic_Exploration/As_above_so_below_lunar_rescue_device_tested_underwater}
\showURL{%
\tempurl}


\bibitem[ESA(2021b)]%
        {NEST}
\bibfield{author}{\bibinfo{person}{ESA}.} \bibinfo{year}{2021}\natexlab{b}.
\newblock \bibinfo{title}{Building a toolkit for the Moon}.
\newblock
\newblock
\urldef\tempurl%
\url{https://www.esa.int/Science_Exploration/Human_and_Robotic_Exploration/Building_a_toolkit_for_the_Moon}
\showURL{%
\tempurl}


\bibitem[ESA(2021c)]%
        {equipment}
\bibfield{author}{\bibinfo{person}{ESA}.} \bibinfo{year}{2021}\natexlab{c}.
\newblock \bibinfo{title}{Moon walking underwater}.
\newblock
\newblock
\urldef\tempurl%
\url{https://www.esa.int/ESA_Multimedia/Images/2019/07/Moon_walking_underwater}
\showURL{%
\tempurl}


\bibitem[Fitzgerald and Ishii(2018)]%
        {fitzgerald2018mediate}
\bibfield{author}{\bibinfo{person}{Daniel Fitzgerald} {and}
  \bibinfo{person}{Hiroshi Ishii}.} \bibinfo{year}{2018}\natexlab{}.
\newblock \showarticletitle{Mediate: A spatial tangible interface for mixed
  reality}. In \bibinfo{booktitle}{\emph{Extended Abstracts of the 2018 CHI
  Conference on Human Factors in Computing Systems}}. \bibinfo{pages}{1--6}.
\newblock


\bibitem[Foust(2021)]%
        {Foust2021}
\bibfield{author}{\bibinfo{person}{Jeff Foust}.}
  \bibinfo{year}{2021}\natexlab{}.
\newblock \bibinfo{title}{Lunar spacesuits won’t be ready in time for 2024
  landing}.
\newblock
\newblock
\urldef\tempurl%
\url{https://spacenews.com/lunar-spacesuits-wont-be-ready-in-time-for-2024-landing/#:~:text=In
  further delays to Artemis.}
\showURL{%
\tempurl}


\bibitem[Fussell and Truong(2021)]%
        {fussell2021using}
\bibfield{author}{\bibinfo{person}{Stephanie~G Fussell} {and}
  \bibinfo{person}{Dothang Truong}.} \bibinfo{year}{2021}\natexlab{}.
\newblock \showarticletitle{Using virtual reality for dynamic learning: an
  extended technology acceptance model}.
\newblock \bibinfo{journal}{\emph{Virtual Reality}} (\bibinfo{year}{2021}),
  \bibinfo{pages}{1--19}.
\newblock


\bibitem[Garassini and Mattei(1994)]%
        {garassini1994evening}
\bibfield{author}{\bibinfo{person}{Stefania Garassini} {and}
  \bibinfo{person}{Maria~Grazia Mattei}.} \bibinfo{year}{1994}\natexlab{}.
\newblock \showarticletitle{Evening Presentation The State of the Art in
  Virtual Reality}.
\newblock \bibinfo{journal}{\emph{Human and Machine Vision}}
  (\bibinfo{year}{1994}), \bibinfo{pages}{385--388}.
\newblock


\bibitem[Gasser et~al\mbox{.}(2019)]%
        {gasser2019}
\bibfield{author}{\bibinfo{person}{Martin Gasser}, \bibinfo{person}{Christa~M.
  Feucht}, \bibinfo{person}{Árni B.~Stefánsson}, {and}
  \bibinfo{person}{Michael~Chalmer Dunn}.} \bibinfo{year}{2019}\natexlab{}.
\newblock \showarticletitle{Lava tube analogue test site, for Moon and Mars in
  West Iceland}.
\newblock \bibinfo{journal}{\emph{EPSC-DPS Joint Meeting 2019, held 15-20
  September 2019 in Geneva, Switzerland, id. EPSC-DPS2019-182}}.
\newblock


\bibitem[Geng et~al\mbox{.}(2017)]%
        {geng2017virtual}
\bibfield{author}{\bibinfo{person}{Jie Geng}, \bibinfo{person}{Ying Li},
  \bibinfo{person}{Ranran Wang}, \bibinfo{person}{Zili Wang},
  \bibinfo{person}{Chuan Lv}, {and} \bibinfo{person}{Dong Zhou}.}
  \bibinfo{year}{2017}\natexlab{}.
\newblock \showarticletitle{A virtual maintenance-based approach for satellite
  assembling and troubleshooting assessment}.
\newblock \bibinfo{journal}{\emph{Acta Astronautica}}  \bibinfo{volume}{138}
  (\bibinfo{year}{2017}), \bibinfo{pages}{434--453}.
\newblock


\bibitem[Goedicke et~al\mbox{.}(2018)]%
        {goedicke2018vr}
\bibfield{author}{\bibinfo{person}{David Goedicke}, \bibinfo{person}{Jamy Li},
  \bibinfo{person}{Vanessa Evers}, {and} \bibinfo{person}{Wendy Ju}.}
  \bibinfo{year}{2018}\natexlab{}.
\newblock \showarticletitle{Vr-oom: Virtual reality on-road driving
  simulation}.
\newblock \bibinfo{journal}{\emph{Proceedings of the 2018 CHI Conference on
  Human Factors in Computing Systems}}, \bibinfo{pages}{1--11}.
\newblock


\bibitem[Gollins et~al\mbox{.}(2020)]%
        {gollins2020}
\bibfield{author}{\bibinfo{person}{Nick Gollins}, \bibinfo{person}{Shahrzad
  Timman}, \bibinfo{person}{Max Braun}, {and} \bibinfo{person}{Markus
  Landgraf}.} \bibinfo{year}{2020}\natexlab{}.
\newblock \showarticletitle{Building a European lunar capability with the
  European large logistic lander}.
\newblock \bibinfo{journal}{\emph{EGU General Assembly Conference}}.
\newblock


\bibitem[Group(2021)]%
        {thales1}
\bibfield{author}{\bibinfo{person}{Thales Group}.}
  \bibinfo{year}{2021}\natexlab{}.
\newblock \bibinfo{title}{ESA Astronauts Alexander Gerst and Luca Parmitano use
  their avatars to check-out future habitation accommodation on the lunar
  Gateway}.
\newblock
\newblock
\urldef\tempurl%
\url{https://www.thalesgroup.com/en/worldwide/space/news/esa-astronauts-alexander-gerst-and-luca-parmitano-use-their-avatars-check-out}
\showURL{%
\tempurl}


\bibitem[He et~al\mbox{.}(2017)]%
        {he2017physhare}
\bibfield{author}{\bibinfo{person}{Zhenyi He}, \bibinfo{person}{Fengyuan Zhu},
  {and} \bibinfo{person}{Ken Perlin}.} \bibinfo{year}{2017}\natexlab{}.
\newblock \showarticletitle{Physhare: Sharing physical interaction in virtual
  reality}.
\newblock \bibinfo{journal}{\emph{Adjunct Publication of the 30th Annual ACM
  Symposium on User Interface Software and Technology}},
  \bibinfo{pages}{17--19}.
\newblock


\bibitem[Helin(2017)]%
        {helin2017augmented}
\bibfield{author}{\bibinfo{person}{Kaj Helin}.}
  \bibinfo{year}{2017}\natexlab{}.
\newblock \showarticletitle{Augmented reality for AIT, AIV and operations}.
\newblock \bibinfo{journal}{\emph{Space Engineering and Technology Final
  Presentation Days}}.
\newblock


\bibitem[Helin et~al\mbox{.}(2018)]%
        {helin2018user}
\bibfield{author}{\bibinfo{person}{Kaj Helin}, \bibinfo{person}{Timo Kuula},
  \bibinfo{person}{Carlo Vizzi}, \bibinfo{person}{Jaakko Karjalainen}, {and}
  \bibinfo{person}{Alla Vovk}.} \bibinfo{year}{2018}\natexlab{}.
\newblock \showarticletitle{User experience of augmented reality system for
  astronaut's manual work support}.
\newblock \bibinfo{journal}{\emph{Frontiers in Robotics and AI}}
  \bibinfo{volume}{5} (\bibinfo{year}{2018}), \bibinfo{pages}{106}.
\newblock


\bibitem[Hubenschmid et~al\mbox{.}(2022)]%
        {hubenschmid2022relive}
\bibfield{author}{\bibinfo{person}{Sebastian Hubenschmid},
  \bibinfo{person}{Jonathan Wieland}, \bibinfo{person}{Daniel~Immanuel Fink},
  \bibinfo{person}{Andrea Batch}, \bibinfo{person}{Johannes Zagermann},
  \bibinfo{person}{Niklas Elmqvist}, {and} \bibinfo{person}{Harald Reiterer}.}
  \bibinfo{year}{2022}\natexlab{}.
\newblock \showarticletitle{ReLive: Bridging In-Situ and Ex-Situ Visual
  Analytics for Analyzing Mixed Reality User Studies}. In
  \bibinfo{booktitle}{\emph{CHI Conference on Human Factors in Computing
  Systems}}. \bibinfo{pages}{1--20}.
\newblock


\bibitem[Jacobson and Rumbaugh(1999)]%
        {jacobson1999}
\bibfield{author}{\bibinfo{person}{Lvar Jacobson} {and} \bibinfo{person}{James
  Grady~Booc Rumbaugh}.} \bibinfo{year}{1999}\natexlab{}.
\newblock \bibinfo{booktitle}{\emph{The UnifiedModeling Language Reference
  Manual}}.
\newblock \bibinfo{publisher}{Addison-Wesley Longman}.
\newblock


\bibitem[Jerald(2015)]%
        {jerald2015vr}
\bibfield{author}{\bibinfo{person}{Jason Jerald}.}
  \bibinfo{year}{2015}\natexlab{}.
\newblock \bibinfo{booktitle}{\emph{The VR book: Human-centered design for
  virtual reality}}.
\newblock \bibinfo{publisher}{Morgan \& Claypool}.
\newblock


\bibitem[Jetter et~al\mbox{.}(2020)]%
        {jetter2020vr}
\bibfield{author}{\bibinfo{person}{Hans-Christian Jetter},
  \bibinfo{person}{Roman R{\"a}dle}, \bibinfo{person}{Tiare Feuchtner},
  \bibinfo{person}{Christoph Anthes}, \bibinfo{person}{Judith Friedl}, {and}
  \bibinfo{person}{Clemens~Nylandsted Klokmose}.}
  \bibinfo{year}{2020}\natexlab{}.
\newblock \showarticletitle{" in vr, everything is possible!": Sketching and
  simulating spatially-aware interactive spaces in virtual reality}. In
  \bibinfo{booktitle}{\emph{Proceedings of the 2020 CHI Conference on Human
  Factors in Computing Systems}}. \bibinfo{pages}{1--16}.
\newblock


\bibitem[Kope{\'c} et~al\mbox{.}(2021)]%
        {kopec2021participatory}
\bibfield{author}{\bibinfo{person}{Wies{\l}aw Kope{\'c}},
  \bibinfo{person}{Cezary Biele}, \bibinfo{person}{Monika Kornacka},
  \bibinfo{person}{Grzegorz Pochwatko}, \bibinfo{person}{Anna Jaskulska},
  \bibinfo{person}{Kinga Skorupska}, \bibinfo{person}{Julia Paluch},
  \bibinfo{person}{Piotr Gago}, \bibinfo{person}{Barbara Karpowicz},
  \bibinfo{person}{Marcin Niewi{\'n}ski}, {et~al\mbox{.}}}
  \bibinfo{year}{2021}\natexlab{}.
\newblock \showarticletitle{Participatory design landscape for the
  human-machine collaboration, interaction and automation at the frontiers of
  HCI (PDL 2021)}. In \bibinfo{booktitle}{\emph{Human-Computer
  Interaction--INTERACT 2021: 18th IFIP TC 13 International Conference, Bari,
  Italy, August 30--September 3, 2021, Proceedings, Part V}}. Springer,
  \bibinfo{pages}{564--569}.
\newblock


\bibitem[Landgraf et~al\mbox{.}(2022)]%
        {landgraf2022autonomous}
\bibfield{author}{\bibinfo{person}{Markus Landgraf}, \bibinfo{person}{Ludovic
  Duvet}, \bibinfo{person}{Alexander Cropp}, \bibinfo{person}{Gustavo Alvarez},
  \bibinfo{person}{Grzegorz Ambroszkiewicz}, \bibinfo{person}{Massimiliano
  Bottacini}, \bibinfo{person}{Pierre Brunner}, \bibinfo{person}{Lorenzo
  Bucci}, \bibinfo{person}{William Carey}, \bibinfo{person}{Andrea
  Emanuele~Maria Casini}, {et~al\mbox{.}}} \bibinfo{year}{2022}\natexlab{}.
\newblock \showarticletitle{Autonomous Access to the Moon for Europe: The
  European Large Logistic Lander}. In \bibinfo{booktitle}{\emph{Proceedings of
  the International Astronautical Congress, IAC}}.
\newblock


\bibitem[Lee et~al\mbox{.}(2019)]%
        {lee2019design}
\bibfield{author}{\bibinfo{person}{Jee~Hyun Lee}, \bibinfo{person}{Eun~Kyoung
  Yang}, {and} \bibinfo{person}{Zhong~Yuan Sun}.}
  \bibinfo{year}{2019}\natexlab{}.
\newblock \showarticletitle{Design cognitive actions stimulating creativity in
  the VR design environment}.
\newblock In \bibinfo{booktitle}{\emph{Proceedings of the 2019 on Creativity
  and Cognition}}. \bibinfo{pages}{604--611}.
\newblock


\bibitem[Lindblad et~al\mbox{.}(2016)]%
        {Lindblad}
\bibfield{author}{\bibinfo{person}{Louise Lindblad}, \bibinfo{person}{Marco
  Witzmann}, {and} \bibinfo{person}{Simon vanden Bussche}.}
  \bibinfo{year}{2016}\natexlab{}.
\newblock \showarticletitle{SYSTEMS ENGINEERING FROM A WEB BROWSER: TURNING
  MBSE INTO INDUSTRIAL REALITY}.
\newblock \bibinfo{journal}{\emph{SECESA}}.
\newblock
\urldef\tempurl%
\url{https://www.valispace.com/wp-content/uploads/2021/02/Lindblad-SECESA-2016-Valispace-web-browser-engineering.pdf}
\showURL{%
\tempurl}


\bibitem[Lindley et~al\mbox{.}(2017)]%
        {Lindley2017}
\bibfield{author}{\bibinfo{person}{Joseph Lindley}, \bibinfo{person}{Paul
  Coulton}, {and} \bibinfo{person}{Miriam Sturdee}.}
  \bibinfo{year}{2017}\natexlab{}.
\newblock \showarticletitle{Implications for Adoption}.
\newblock \bibinfo{journal}{\emph{Proceedings of the 2017 CHI Conference on
  Human Factors in Computing Systems}}, \bibinfo{pages}{265--277}.
\newblock
\showISBNx{9781450346559}
\urldef\tempurl%
\url{https://doi.org/10.1145/3025453.3025742}
\showDOI{\tempurl}


\bibitem[Loftin and Kenney(1995)]%
        {loftin1995training}
\bibfield{author}{\bibinfo{person}{R~Bowen Loftin} {and} \bibinfo{person}{P
  Kenney}.} \bibinfo{year}{1995}\natexlab{}.
\newblock \showarticletitle{Training the Hubble space telescope flight team}.
\newblock \bibinfo{journal}{\emph{IEEE Computer Graphics and Applications}}
  \bibinfo{volume}{15}, \bibinfo{number}{5} (\bibinfo{year}{1995}),
  \bibinfo{pages}{31--37}.
\newblock


\bibitem[Lohse et~al\mbox{.}(2014)]%
        {lohse2014virtual}
\bibfield{author}{\bibinfo{person}{Keith~R Lohse}, \bibinfo{person}{Courtney
  G~E Hilderman}, \bibinfo{person}{Katharine~L Cheung}, \bibinfo{person}{Sandy
  Tatla}, {and} \bibinfo{person}{H~F~Machiel der Loos}.}
  \bibinfo{year}{2014}\natexlab{}.
\newblock \showarticletitle{Virtual reality therapy for adults post-stroke: a
  systematic review and meta-analysis exploring virtual environments and
  commercial games in therapy}.
\newblock \bibinfo{journal}{\emph{PloS one}}  \bibinfo{volume}{9}
  (\bibinfo{year}{2014}), \bibinfo{pages}{e93318}.
\newblock
Issue 3.


\bibitem[Léveillé(2010)]%
        {Leveille2010}
\bibfield{author}{\bibinfo{person}{Richard Léveillé}.}
  \bibinfo{year}{2010}\natexlab{}.
\newblock \showarticletitle{A half-century of terrestrial analog studies: From
  craters on the Moon to searching for life on Mars}.
\newblock \bibinfo{journal}{\emph{Planetary and Space Science}}
  \bibinfo{volume}{58} (\bibinfo{date}{3} \bibinfo{year}{2010}),
  \bibinfo{pages}{631--638}.
\newblock
Issue 4.
\showISSN{0032-0633}
\urldef\tempurl%
\url{https://doi.org/10.1016/J.PSS.2009.04.001}
\showDOI{\tempurl}


\bibitem[M{\"a}kel{\"a} et~al\mbox{.}(2020)]%
        {makela2020virtual}
\bibfield{author}{\bibinfo{person}{Ville M{\"a}kel{\"a}}, \bibinfo{person}{Rivu
  Radiah}, \bibinfo{person}{Saleh Alsherif}, \bibinfo{person}{Mohamed Khamis},
  \bibinfo{person}{Chong Xiao}, \bibinfo{person}{Lisa Borchert},
  \bibinfo{person}{Albrecht Schmidt}, {and} \bibinfo{person}{Florian Alt}.}
  \bibinfo{year}{2020}\natexlab{}.
\newblock \showarticletitle{Virtual field studies: conducting studies on public
  displays in virtual reality}. In \bibinfo{booktitle}{\emph{Proceedings of the
  2020 CHI Conference on Human Factors in Computing Systems}}.
  \bibinfo{pages}{1--15}.
\newblock


\bibitem[McVeigh-Schultz et~al\mbox{.}(2018)]%
        {mcveigh2018immersive}
\bibfield{author}{\bibinfo{person}{Joshua McVeigh-Schultz},
  \bibinfo{person}{Max Kreminski}, \bibinfo{person}{Keshav Prasad},
  \bibinfo{person}{Perry Hoberman}, {and} \bibinfo{person}{Scott~S Fisher}.}
  \bibinfo{year}{2018}\natexlab{}.
\newblock \showarticletitle{Immersive design fiction: Using VR to prototype
  speculative interfaces and interaction rituals within a virtual storyworld}.
  In \bibinfo{booktitle}{\emph{Proceedings of the 2018 designing interactive
  systems conference}}. \bibinfo{pages}{817--829}.
\newblock


\bibitem[Miner and Stansfield(1994)]%
        {miner1994interactive}
\bibfield{author}{\bibinfo{person}{Nadine~E Miner} {and}
  \bibinfo{person}{Sharon~A Stansfield}.} \bibinfo{year}{1994}\natexlab{}.
\newblock \showarticletitle{An interactive virtual reality simulation system
  for robot control and operator training}.
\newblock \bibinfo{journal}{\emph{Proceedings of the 1994 IEEE International
  Conference on Robotics and Automation}}, \bibinfo{pages}{1428--1435}.
\newblock


\bibitem[Moore(1989)]%
        {Moore1989}
\bibfield{author}{\bibinfo{person}{Thomas~P Moore}.}
  \bibinfo{year}{1989}\natexlab{}.
\newblock \showarticletitle{US space flight experience. Physical exertion and
  metabolic demand of extravehicular activity: Past, present, and future}. In
  \bibinfo{booktitle}{\emph{NASA, Johnson Space Center, Workshop on Exercise
  Prescription for Long-Duration Space Flight}}.
\newblock


\bibitem[Nilsson et~al\mbox{.}(2022)]%
        {nilsson2022using}
\bibfield{author}{\bibinfo{person}{Tommy Nilsson}, \bibinfo{person}{Flavie
  Rometsch}, \bibinfo{person}{Andrea Emanuele~Maria Casini},
  \bibinfo{person}{Enrico Guerra}, \bibinfo{person}{Leonie Becker},
  \bibinfo{person}{Andreas Treuer}, \bibinfo{person}{Paul de Medeiros},
  \bibinfo{person}{Hanjo Schnellbaecher}, \bibinfo{person}{Anna Vock}, {and}
  \bibinfo{person}{Aidan Cowley}.} \bibinfo{year}{2022}\natexlab{}.
\newblock \showarticletitle{Using Virtual Reality to Design and Evaluate a
  Lunar Lander: The EL3 Case Study}. In \bibinfo{booktitle}{\emph{CHI
  Conference on Human Factors in Computing Systems Extended Abstracts}}.
  \bibinfo{pages}{1--7}.
\newblock


\bibitem[Obrist et~al\mbox{.}(2019)]%
        {obrist2019space}
\bibfield{author}{\bibinfo{person}{Marianna Obrist}, \bibinfo{person}{Yunwen
  Tu}, \bibinfo{person}{Lining Yao}, {and} \bibinfo{person}{Carlos Velasco}.}
  \bibinfo{year}{2019}\natexlab{}.
\newblock \showarticletitle{Space food experiences: designing passenger's
  eating experiences for future space travel scenarios}.
\newblock \bibinfo{journal}{\emph{Frontiers in Computer Science}}
  \bibinfo{volume}{1} (\bibinfo{year}{2019}), \bibinfo{pages}{3}.
\newblock


\bibitem[Orr et~al\mbox{.}(2022)]%
        {orr2022effects}
\bibfield{author}{\bibinfo{person}{Sophie Orr}, \bibinfo{person}{James Casler},
  \bibinfo{person}{Jesse Rhoades}, {and} \bibinfo{person}{Pablo de Le{\'o}n}.}
  \bibinfo{year}{2022}\natexlab{}.
\newblock \showarticletitle{Effects of walking, running, and skipping under
  simulated reduced gravity using the NASA Active Response Gravity Offload
  System (ARGOS)}.
\newblock \bibinfo{journal}{\emph{Acta Astronautica}} (\bibinfo{year}{2022}).
\newblock


\bibitem[Osinski et~al\mbox{.}(2006)]%
        {Osinski2006}
\bibfield{author}{\bibinfo{person}{Gordon~R. Osinski}, \bibinfo{person}{Richard
  Léveillé}, \bibinfo{person}{Alain Berinstain}, \bibinfo{person}{Martin
  Lebeuf}, {and} \bibinfo{person}{Matthew Bamsey}.}
  \bibinfo{year}{2006}\natexlab{}.
\newblock \showarticletitle{Terrestrial analogues to Mars and the Moon:
  Canada’s role}.
\newblock \bibinfo{journal}{\emph{Geoscience Canada}}  \bibinfo{volume}{33}
  (\bibinfo{year}{2006}), \bibinfo{pages}{175--188}.
\newblock
Issue 4.


\bibitem[Osterlund and Lawrence(2012)]%
        {osterlund2011}
\bibfield{author}{\bibinfo{person}{Jeffrey Osterlund} {and}
  \bibinfo{person}{Brad Lawrence}.} \bibinfo{year}{2012}\natexlab{}.
\newblock \showarticletitle{Virtual reality: Avatars in human spaceflight
  training}.
\newblock \bibinfo{journal}{\emph{Acta Astronautica}}  \bibinfo{volume}{71}
  (\bibinfo{year}{2012}), \bibinfo{pages}{139--150}.
\newblock


\bibitem[Pataranutaporn et~al\mbox{.}(2021)]%
        {pataranutaporn2021spacechi}
\bibfield{author}{\bibinfo{person}{Pat Pataranutaporn},
  \bibinfo{person}{Valentina Sumini}, \bibinfo{person}{Ariel Ekblaw},
  \bibinfo{person}{Melodie Yashar}, \bibinfo{person}{Sandra
  H{\"a}uplik-Meusburger}, \bibinfo{person}{Susanna Testa},
  \bibinfo{person}{Marianna Obrist}, \bibinfo{person}{Dorit Donoviel},
  \bibinfo{person}{Joseph Paradiso}, {and} \bibinfo{person}{Pattie Maes}.}
  \bibinfo{year}{2021}\natexlab{}.
\newblock \showarticletitle{SpaceCHI: Designing Human-Computer Interaction
  Systems for Space Exploration}. In \bibinfo{booktitle}{\emph{Extended
  Abstracts of the 2021 CHI Conference on Human Factors in Computing Systems}}.
  \bibinfo{pages}{1--6}.
\newblock


\bibitem[Pataranutaporn et~al\mbox{.}(2022)]%
        {pataranutaporn2022spacechi}
\bibfield{author}{\bibinfo{person}{Pat Pataranutaporn},
  \bibinfo{person}{Valentina Sumini}, \bibinfo{person}{Melodie Yashar},
  \bibinfo{person}{Susanna Testa}, \bibinfo{person}{Marianna Obrist},
  \bibinfo{person}{Scott Davidoff}, \bibinfo{person}{Amber~M Paul},
  \bibinfo{person}{Dorit Donoviel}, \bibinfo{person}{Jimmy Wu},
  \bibinfo{person}{Sands~A Fish}, {et~al\mbox{.}}}
  \bibinfo{year}{2022}\natexlab{}.
\newblock \showarticletitle{SpaceCHI 2.0: Advancing Human-Computer Interaction
  Systems for Space Exploration}. In \bibinfo{booktitle}{\emph{CHI Conference
  on Human Factors in Computing Systems Extended Abstracts}}.
  \bibinfo{pages}{1--7}.
\newblock


\bibitem[Platt et~al\mbox{.}(2014)]%
        {platt2014participatory}
\bibfield{author}{\bibinfo{person}{Donald Platt}, \bibinfo{person}{Patrick
  Millot}, {and} \bibinfo{person}{Guy~Andre Boy}.}
  \bibinfo{year}{2014}\natexlab{}.
\newblock \showarticletitle{Participatory design of a cooperative exploration
  mediation tool for human deep space risk mitigation}. In
  \bibinfo{booktitle}{\emph{International Conference on Engineering Psychology
  and Cognitive Ergonomics}}. Springer, \bibinfo{pages}{363--374}.
\newblock


\bibitem[Porter et~al\mbox{.}(2020)]%
        {porter2020soft}
\bibfield{author}{\bibinfo{person}{Allison~P Porter}, \bibinfo{person}{Barnaba
  Marchesini}, \bibinfo{person}{Irina Potryasilova}, \bibinfo{person}{Enrico
  Rossetto}, {and} \bibinfo{person}{Dava~J Newman}.}
  \bibinfo{year}{2020}\natexlab{}.
\newblock \showarticletitle{Soft exoskeleton knee prototype for advanced space
  suits and planetary exploration}. In \bibinfo{booktitle}{\emph{2020 IEEE
  Aerospace Conference}}. IEEE, \bibinfo{pages}{1--13}.
\newblock


\bibitem[Reinschluessel et~al\mbox{.}(2017)]%
        {reinschluessel2017virtual}
\bibfield{author}{\bibinfo{person}{Anke~Verena Reinschluessel},
  \bibinfo{person}{Joern Teuber}, \bibinfo{person}{Marc Herrlich},
  \bibinfo{person}{Jeffrey Bissel}, \bibinfo{person}{Melanie van Eikeren},
  \bibinfo{person}{Johannes Ganser}, \bibinfo{person}{Felicia Koeller},
  \bibinfo{person}{Fenja Kollasch}, \bibinfo{person}{Thomas Mildner},
  \bibinfo{person}{Luca Raimondo}, {et~al\mbox{.}}}
  \bibinfo{year}{2017}\natexlab{}.
\newblock \showarticletitle{Virtual reality for user-centered design and
  evaluation of touch-free interaction techniques for navigating medical images
  in the operating room}.
\newblock \bibinfo{journal}{\emph{Proceedings of the 2017 CHI Conference
  Extended Abstracts on Human Factors in Computing Systems}},
  \bibinfo{pages}{2001--2009}.
\newblock


\bibitem[Rhodes(2008)]%
        {rhodes2008addressing}
\bibfield{author}{\bibinfo{person}{D Rhodes}.} \bibinfo{year}{2008}\natexlab{}.
\newblock \showarticletitle{Addressing systems engineering challenges through
  collaborative research}.
\newblock \bibinfo{journal}{\emph{SEARI—Systems engineering advancement
  research initiative}} (\bibinfo{year}{2008}).
\newblock


\bibitem[Rometsch et~al\mbox{.}(2022)]%
        {rometsch2022towards}
\bibfield{author}{\bibinfo{person}{Flavie Rometsch}, \bibinfo{person}{Tommy
  Nilsson}, \bibinfo{person}{Paul de Medeiros}, \bibinfo{person}{Andreas
  Treuer}, \bibinfo{person}{Aidan Cowley}, \bibinfo{person}{Andrea
  Emanuele~Maria Casini}, \bibinfo{person}{Ludovic Duvet},
  \bibinfo{person}{Anna Vock}, \bibinfo{person}{Leonie Becker},
  \bibinfo{person}{Hanjo Schnellb{\"a}cher}, {et~al\mbox{.}}}
  \bibinfo{year}{2022}\natexlab{}.
\newblock \showarticletitle{Towards a human-centred framework for
  conceptualization of lunar surface solutions}. In
  \bibinfo{booktitle}{\emph{Proceedings of the International Astronautical
  Congress, IAC}}.
\newblock


\bibitem[Ross et~al\mbox{.}(2018)]%
        {ross2018nasa}
\bibfield{author}{\bibinfo{person}{Amy Ross}, \bibinfo{person}{Richard Rhodes},
  {and} \bibinfo{person}{Shane McFarland}.} \bibinfo{year}{2018}\natexlab{}.
\newblock \showarticletitle{NASA’s advanced extra-vehicular activity space
  suit pressure garment 2018 status and development plan}.
\newblock


\bibitem[Salovaara et~al\mbox{.}(2017)]%
        {Salovaara2017}
\bibfield{author}{\bibinfo{person}{Antti Salovaara}, \bibinfo{person}{Antti
  Oulasvirta}, {and} \bibinfo{person}{Giulio Jacucci}.}
  \bibinfo{year}{2017}\natexlab{}.
\newblock \showarticletitle{Evaluation of Prototypes and the Problem of
  Possible Futures}.
\newblock \bibinfo{journal}{\emph{Proceedings of the 2017 CHI Conference on
  Human Factors in Computing Systems}}, \bibinfo{pages}{2064--2077}.
\newblock
\showISBNx{9781450346559}
\urldef\tempurl%
\url{https://doi.org/10.1145/3025453.3025658}
\showDOI{\tempurl}


\bibitem[Sauro et~al\mbox{.}(2018)]%
        {Pangaea-X}
\bibfield{author}{\bibinfo{person}{Francesco Sauro}, \bibinfo{person}{M
  Massironi}, \bibinfo{person}{Riccardo Pozzobon}, \bibinfo{person}{H
  Hiesinger}, \bibinfo{person}{Nicolas Mangold}, \bibinfo{person}{Jesús
  Martínez-Frías}, \bibinfo{person}{Charles Cockell}, {and}
  \bibinfo{person}{Loredana Bessone}.} \bibinfo{year}{2018}\natexlab{}.
\newblock \showarticletitle{The ESA PANGAEA and PANGAEA eXtension testing
  analogue}.
\newblock
\urldef\tempurl%
\url{https://www.researchgate.net/project/PANGAEA-Planetary-ANalogue-Geological-and-Astrobiological-Exercise-for-Astronauts}
\showURL{%
\tempurl}


\bibitem[Sauro et~al\mbox{.}(2023)]%
        {sauro2023training}
\bibfield{author}{\bibinfo{person}{Francesco Sauro}, \bibinfo{person}{Samuel~J
  Payler}, \bibinfo{person}{Matteo Massironi}, \bibinfo{person}{Riccardo
  Pozzobon}, \bibinfo{person}{Harald Hiesinger}, \bibinfo{person}{Nicolas
  Mangold}, \bibinfo{person}{Charles~S Cockell},
  \bibinfo{person}{Jesus~Mart{\'\i}nez Frias}, \bibinfo{person}{K{\aa}re
  Kullerud}, \bibinfo{person}{Leonardo Turchi}, {et~al\mbox{.}}}
  \bibinfo{year}{2023}\natexlab{}.
\newblock \showarticletitle{Training astronauts for scientific exploration on
  planetary surfaces: The ESA PANGAEA programme}.
\newblock \bibinfo{journal}{\emph{Acta Astronautica}}  \bibinfo{volume}{204}
  (\bibinfo{year}{2023}), \bibinfo{pages}{222--238}.
\newblock


\bibitem[Seedhouse(2022)]%
        {seedhouse2022starship}
\bibfield{author}{\bibinfo{person}{Erik Seedhouse}.}
  \bibinfo{year}{2022}\natexlab{}.
\newblock \showarticletitle{Starship}.
\newblock In \bibinfo{booktitle}{\emph{SpaceX}}. \bibinfo{publisher}{Springer},
  \bibinfo{pages}{171--188}.
\newblock


\bibitem[Shaw et~al\mbox{.}(2019)]%
        {shaw2019heat}
\bibfield{author}{\bibinfo{person}{Emily Shaw}, \bibinfo{person}{Tessa Roper},
  \bibinfo{person}{Tommy Nilsson}, \bibinfo{person}{Glyn Lawson},
  \bibinfo{person}{Sue V~G Cobb}, {and} \bibinfo{person}{Daniel Miller}.}
  \bibinfo{year}{2019}\natexlab{}.
\newblock \showarticletitle{The heat is on: Exploring user behaviour in a
  multisensory virtual environment for fire evacuation}.
\newblock \bibinfo{journal}{\emph{Proceedings of the 2019 CHI Conference on
  Human Factors in Computing Systems}}, \bibinfo{pages}{1--13}.
\newblock


\bibitem[Sheridan(1992)]%
        {sheridan1992telerobotics}
\bibfield{author}{\bibinfo{person}{Thomas~B Sheridan}.}
  \bibinfo{year}{1992}\natexlab{}.
\newblock \bibinfo{booktitle}{\emph{Telerobotics, automation, and human
  supervisory control}}.
\newblock \bibinfo{publisher}{MIT press}.
\newblock


\bibitem[Simeone et~al\mbox{.}(2022)]%
        {simeone2022immersive}
\bibfield{author}{\bibinfo{person}{Adalberto~L Simeone}, \bibinfo{person}{Robbe
  Cools}, \bibinfo{person}{Stan Depuydt}, \bibinfo{person}{Jo{\~a}o~Maria
  Gomes}, \bibinfo{person}{Piet Goris}, \bibinfo{person}{Joseph Grocott},
  \bibinfo{person}{Augusto Esteves}, {and} \bibinfo{person}{Kathrin Gerling}.}
  \bibinfo{year}{2022}\natexlab{}.
\newblock \showarticletitle{Immersive Speculative Enactments: Bringing Future
  Scenarios and Technology to Life Using Virtual Reality}. In
  \bibinfo{booktitle}{\emph{CHI Conference on Human Factors in Computing
  Systems}}. \bibinfo{pages}{1--20}.
\newblock


\bibitem[Smith et~al\mbox{.}(2017)]%
        {smith2017summary}
\bibfield{author}{\bibinfo{person}{David~E Smith}, \bibinfo{person}{Maria~T
  Zuber}, \bibinfo{person}{Gregory~A Neumann}, \bibinfo{person}{Erwan
  Mazarico}, \bibinfo{person}{Frank~G Lemoine}, \bibinfo{person}{James W~Head
  III}, \bibinfo{person}{Paul~G Lucey}, \bibinfo{person}{Oded Aharonson},
  \bibinfo{person}{Mark~S Robinson}, \bibinfo{person}{Xiaoli Sun},
  {et~al\mbox{.}}} \bibinfo{year}{2017}\natexlab{}.
\newblock \showarticletitle{Summary of the results from the lunar orbiter laser
  altimeter after seven years in lunar orbit}.
\newblock \bibinfo{journal}{\emph{Icarus}}  \bibinfo{volume}{283}
  (\bibinfo{year}{2017}), \bibinfo{pages}{70--91}.
\newblock


\bibitem[Smith et~al\mbox{.}(2020)]%
        {Smith2020}
\bibfield{author}{\bibinfo{person}{Marshall Smith}, \bibinfo{person}{Douglas
  Craig}, \bibinfo{person}{Nicole Herrmann}, \bibinfo{person}{Erin Mahoney},
  \bibinfo{person}{Jonathan Krezel}, \bibinfo{person}{Nate McIntyre}, {and}
  \bibinfo{person}{Kandyce Goodliff}.} \bibinfo{year}{2020}\natexlab{}.
\newblock \showarticletitle{The Artemis Program: An Overview of NASA's
  Activities to Return Humans to the Moon}.
\newblock \bibinfo{journal}{\emph{2020 IEEE Aerospace Conference}},
  \bibinfo{pages}{1--10}.
\newblock
\showISBNx{978-1-7281-2734-7}
\urldef\tempurl%
\url{https://doi.org/10.1109/AERO47225.2020.9172323}
\showDOI{\tempurl}


\bibitem[Somin et~al\mbox{.}(2021)]%
        {somin2021breachmob}
\bibfield{author}{\bibinfo{person}{Lior Somin}, \bibinfo{person}{Zachary
  McKendrick}, \bibinfo{person}{Patrick Finn}, {and} \bibinfo{person}{Ehud
  Sharlin}.} \bibinfo{year}{2021}\natexlab{}.
\newblock \showarticletitle{BreachMob: Detecting Vulnerabilities in Physical
  Environments Using Virtual Reality}.
\newblock \bibinfo{journal}{\emph{Proceedings of the 27th ACM Symposium on
  Virtual Reality Software and Technology}}, \bibinfo{pages}{1--6}.
\newblock


\bibitem[Sportillo et~al\mbox{.}(2017)]%
        {sportillo2017immersive}
\bibfield{author}{\bibinfo{person}{Daniele Sportillo}, \bibinfo{person}{Alexis
  Paljic}, \bibinfo{person}{Mehdi Boukhris}, \bibinfo{person}{Philippe Fuchs},
  \bibinfo{person}{Luciano Ojeda}, {and} \bibinfo{person}{Vincent Roussarie}.}
  \bibinfo{year}{2017}\natexlab{}.
\newblock \showarticletitle{An immersive Virtual Reality system for
  semi-autonomous driving simulation: a comparison between realistic and 6-DoF
  controller-based interaction}.
\newblock \bibinfo{journal}{\emph{Proceedings of the 9th International
  Conference on Computer and Automation Engineering}}, \bibinfo{pages}{6--10}.
\newblock


\bibitem[Sumini et~al\mbox{.}(2020)]%
        {sumini2020spacehuman}
\bibfield{author}{\bibinfo{person}{Valentina Sumini}, \bibinfo{person}{Manuel
  Muccillo}, \bibinfo{person}{Jamie Milliken}, \bibinfo{person}{Ariel Ekblaw},
  {and} \bibinfo{person}{Joseph Paradiso}.} \bibinfo{year}{2020}\natexlab{}.
\newblock \showarticletitle{SpaceHuman: A Soft Robotic Prosthetic for Space
  Exploration}. In \bibinfo{booktitle}{\emph{Extended Abstracts of the 2020 CHI
  Conference on Human Factors in Computing Systems}}. \bibinfo{pages}{1--8}.
\newblock


\bibitem[Trotta et~al\mbox{.}(2020)]%
        {trotta2020communicating}
\bibfield{author}{\bibinfo{person}{Roberto Trotta}, \bibinfo{person}{Daniel
  Hajas}, \bibinfo{person}{Jos{\'e}~Eliel Camargo-Molina},
  \bibinfo{person}{Robert Cobden}, \bibinfo{person}{Emanuela Maggioni},
  \bibinfo{person}{Marianna Obrist}, {et~al\mbox{.}}}
  \bibinfo{year}{2020}\natexlab{}.
\newblock \showarticletitle{Communicating cosmology with multisensory
  metaphorical experiences}.
\newblock  (\bibinfo{year}{2020}).
\newblock


\bibitem[Turchi et~al\mbox{.}(2021)]%
        {Turchi2021}
\bibfield{author}{\bibinfo{person}{Leonardo Turchi}, \bibinfo{person}{S.~J.
  Payler}, \bibinfo{person}{Francesco Sauro}, \bibinfo{person}{Riccardo
  Pozzobon}, \bibinfo{person}{Matteo Massironi}, {and}
  \bibinfo{person}{Loredana Bessone}.} \bibinfo{year}{2021}\natexlab{}.
\newblock \showarticletitle{The Electronic FieldBook: A system for supporting
  distributed field science operations during astronaut training and human
  planetary exploration}.
\newblock \bibinfo{journal}{\emph{Planetary and Space Science}}
  \bibinfo{volume}{197} (\bibinfo{date}{3} \bibinfo{year}{2021}).
\newblock
\urldef\tempurl%
\url{https://doi.org/10.1016/j.pss.2021.105164}
\showDOI{\tempurl}


\bibitem[Vanoutryve et~al\mbox{.}(2010)]%
        {vanoutryve2010analysis}
\bibfield{author}{\bibinfo{person}{Benjamin Vanoutryve},
  \bibinfo{person}{Diego~De Rosa}, \bibinfo{person}{Richard Fisackerly},
  \bibinfo{person}{Berengere Houdou}, \bibinfo{person}{James Carpenter},
  \bibinfo{person}{Christian Philippe}, \bibinfo{person}{Alain Pradier},
  \bibinfo{person}{Aliac Jojaghaian}, \bibinfo{person}{Sylvie Espinasse}, {and}
  \bibinfo{person}{Bruno Gardini}.} \bibinfo{year}{2010}\natexlab{}.
\newblock \showarticletitle{An analysis of illumination and communication
  conditions near lunar south pole based on Kaguya Data}.
\newblock \bibinfo{journal}{\emph{Proceedings of International Planetary Probe
  Workshop, Barcelona}}.
\newblock


\bibitem[Vock and Nilsson(2022)]%
        {vock2022holistic}
\bibfield{author}{\bibinfo{person}{Anna Vock} {and} \bibinfo{person}{Tommy
  Nilsson}.} \bibinfo{year}{2022}\natexlab{}.
\newblock \showarticletitle{Holistic Outpost Design for Lunar Lava Tubes}. In
  \bibinfo{booktitle}{\emph{Proceedings of the International Astronautical
  Congress, IAC}}.
\newblock


\bibitem[Walch et~al\mbox{.}(2017)]%
        {walch2017evaluating}
\bibfield{author}{\bibinfo{person}{Marcel Walch}, \bibinfo{person}{Julian
  Frommel}, \bibinfo{person}{Katja Rogers}, \bibinfo{person}{Felix Schüssel},
  \bibinfo{person}{Philipp Hock}, \bibinfo{person}{David Dobbelstein}, {and}
  \bibinfo{person}{Michael Weber}.} \bibinfo{year}{2017}\natexlab{}.
\newblock \showarticletitle{Evaluating VR driving simulation from a player
  experience perspective}.
\newblock \bibinfo{journal}{\emph{Proceedings of the 2017 CHI Conference
  Extended Abstracts on Human Factors in Computing Systems}},
  \bibinfo{pages}{2982--2989}.
\newblock


\bibitem[Weber et~al\mbox{.}(2021)]%
        {Weber2021}
\bibfield{author}{\bibinfo{person}{Renee Weber}, \bibinfo{person}{Barbara
  Cohen}, {and} \bibinfo{person}{Samuel Lawrence}.}
  \bibinfo{year}{2021}\natexlab{}.
\newblock \bibinfo{title}{The Artemis III Science Definition Team Report}.
\newblock
\newblock
\urldef\tempurl%
\url{https://www.nasa.gov/sites/default/files/atoms/files/artemis-iii-science-definition-report-12042020c.pdf}
\showURL{%
\tempurl}


\bibitem[Weinzierl(2018)]%
        {weinzierl2018space}
\bibfield{author}{\bibinfo{person}{Matthew Weinzierl}.}
  \bibinfo{year}{2018}\natexlab{}.
\newblock \showarticletitle{Space, the final economic frontier}.
\newblock \bibinfo{journal}{\emph{Journal of Economic Perspectives}}
  \bibinfo{volume}{32}, \bibinfo{number}{2} (\bibinfo{year}{2018}),
  \bibinfo{pages}{173--92}.
\newblock


\bibitem[Whitehouse et~al\mbox{.}(2021)]%
        {ESA_MBSE2021}
\bibfield{author}{\bibinfo{person}{Jamie Whitehouse}, \bibinfo{person}{Alberto
  Gonz{\'a}lez~Fern{\'a}ndez}, \bibinfo{person}{Elaheh Maleki}, {and}
  \bibinfo{person}{Anh~Toan Bui~Long}.} \bibinfo{year}{2021}\natexlab{}.
\newblock \showarticletitle{MBSE at ESA: State of MBSE in ESA Missions and
  Activities}. In \bibinfo{booktitle}{\emph{MBSE 2021 Conference,
  https://indico. esa. int/event/386/timetable}}.
\newblock


\bibitem[Wienrich et~al\mbox{.}(2018)]%
        {wienrich2018assessing}
\bibfield{author}{\bibinfo{person}{Carolin Wienrich}, \bibinfo{person}{Nina
  Döllinger}, \bibinfo{person}{Simon Kock}, \bibinfo{person}{Kristina
  Schindler}, {and} \bibinfo{person}{Ole Traupe}.}
  \bibinfo{year}{2018}\natexlab{}.
\newblock \showarticletitle{Assessing user experience in virtual reality--a
  comparison of different measurements}.
\newblock \bibinfo{journal}{\emph{International Conference of Design, User
  Experience, and Usability}}, \bibinfo{pages}{573--589}.
\newblock


\end{thebibliography}


\appendix

\end{document}